\documentclass[final,3p,times]{elsarticle}

\usepackage{amssymb}

\usepackage{amssymb}
\usepackage[]{amsmath}
\usepackage{graphics}
\usepackage{amssymb}
\usepackage[]{amsmath}
\usepackage{amsthm}
\usepackage{setspace}
\usepackage{epsfig}
\usepackage{subfigure}
\usepackage{extarrows}

\def\[{\begin{equation}}
\def\]{\end{equation}}
\biboptions{numbers,sort&compress}
\usepackage[dvipdfm,colorlinks,linkcolor=red,anchorcolor=blue,citecolor=green]{hyperref}

\usepackage{amsmath}
\usepackage{times}
\usepackage{anysize}
\marginsize{2cm}{2cm}{0cm}{1.6cm}
\usepackage{mathpazo,color}

\newtheorem{lem}{Lemma}
\newtheorem{prop}{Proposition}
\newtheorem{theorem}{Theorem}

\selectfont

\journal{******}

\begin{document}
\begin{frontmatter}



\title{Rogue waves in the massive Thirring model}

\author{Junchao Chen  \fnref{label1}  }
\author{Bo Yang \fnref{label2} }
\author{Bao-Feng Feng \fnref{label3} \corref{cor1}}
\ead{baofeng.feng@utrgv.edu}

\cortext[cor1]{School of Mathematical and Statistical Sciences, The University of Texas Rio Grande Valley, Edinburg, TX 78541, USA.}

\address[label1]{Department of Mathematics, Lishui University, Lishui, 323000, China}

\address[label2]{School of Mathematics and Statistics, Ningbo University, Ningbo 315211, China}

\address[label3]{School of Mathematical and Statistical Sciences, The University of Texas Rio Grande Valley, Edinburg, TX 78541, USA.}

\begin{abstract}
In this paper, general rogue wave solutions in the massive Thirring (MT) model are derived by using
the Kadomtsev-Petviashvili (KP) hierarchy reduction method and these rational solutions are presented explicitly in terms of determinants whose matrix elements are elementary Schur polynomials.
In the reduction process, three reduction conditions including one index- and two dimension-ones are proved to be consistent by only one constraint relation on parameters of tau-functions of the KP-Toda hierarchy.
It is found that the rogue wave solutions in the MT model depend on two background parameters, which influence their orientation and duration.
Differing from many other coupled integrable systems, the MT model only admits the rogue waves of bright-type, and the higher-order rogue waves represent the superposition of fundamental ones in which the non-reducible parameters determine the arrangement patterns of fundamental rogue waves.
Particularly, the super rogue wave at each order can be achieved simply by setting all internal parameters to be zero, resulting in the amplitude of the sole huge peak of order $N$ being $2N+1$ times the background.
Finally, rogue wave patterns are discussed when one of the internal parameters is large. Similar to other integrable equations, the patterns are shown to be associated with the root structures of the Yablonskii-Vorob'ev polynomial hierarchy through a linear transformation.
\end{abstract}
\end{frontmatter}

\section{Introduction}
In recent years, rogue waves appearing in various complex systems have become a fascinating subject of experimental and theoretical studies \cite{onorato2013rogue,kharif2009rogue,solli2007optical,dudley2019rogue,chabchoub2011rogue,bludov2009matter,yan2010financial,bailung2011observation}. They correspond to large-amplitude and spontaneous local excitations with the instability and unpredictability, and hence could cause maritime disasters in oceanography and induce pulse destroys in optics \cite{onorato2013rogue,kharif2009rogue,solli2007optical,dudley2019rogue}.
From the mathematical description, Peregrine soliton characterized by a kind of rational solutions for the focusing nonlinear Schr\"{o}dinger (NLS) equation was discovered to act as the prototype of realistic rogue waves \cite{peregrine1983water}, since it exhibits the local wave structure in temporal-spatial plane and the height of maximum peak at the center reaches to three times the finite background.
Because of the underlying integrability, it was found that Peregrine soliton can be extended to higher-order exact rogue wave solutions in many nonlinear wave systems such as the NLS equation \cite{Akhmediev2009rogue,guo2012nolinear,Dubard2013multi,Ohta2012NLS,Miller2019robust}, the derivative NLS equation \cite{xu2011darboux,guo2013high,chen2019super,BYang2020DNLS}, the Sasa-Satsuma equation \cite{Bandelow2012Sasa,chen2012twisted,mu2016dynamic}, the Manakov equations \cite{guo2011roguecpl,baronio2012solutions,baronio2014vector},
the long-wave-short-wave equations \cite{chen2014dark,chow2013rogue,chen2018jpsj,chen2019pre},
the three-wave resonant interaction equations \cite{Degasperis2013rational,wang2015higher,chen2013watch,zhang2018three,BYang2021TW},
the Davey-Stewartson equations \cite{OhtaJKY2012,OhtaJKY2013} and many others \cite{LLMFZ2016,zhao2013roguewave,wen2015modulational,OhtaJKY2014,yang2020boussinesq,wang2021maxwell,Degasperis2015Bragg,guo2017rogueMT,ye2021super}.
These analytic rational solutions with the higher-order polynomials also represent the localized structure in both space and time coordinates, and could possess multiple intensity peaks or higher peak amplitudes.
In particular, differing from the scalar system, the coupled and vector integrable systems with the additional degrees of freedom could allow the novel counterpart of rogue wave such as dark and four-petaled types.
In addition, more interesting features have been discussed attributed to the explicit expressions for general rogue waves \cite{Akhmediev2009rogue,Kedziora2013classifying,Ankiewicz2017multi,BYang2021patterns1,BYang2021patterns2}.
For instance, a $N$-th order rogue wave usually contains $N(N+1)/2$ elementary (Peregrine) rogue waves \cite{Ankiewicz2017multi},
a super $N$-th order rogue wave that all peaks converge to be a sole huge one has the maximum amplitude $2N+1$ times its background \cite{Akhmediev2009rogue}. Furthermore, the geometric patterns of elementary rogue waves arrangements have been found to be closely related with the root structures of Yablonskii-Vorob'ev polynomials \cite{BYang2021patterns1,BYang2021patterns2}.

The massive Thirring (MT) model with the two-component couplings, proposed in the context of quantum field theory \cite{Thirring1958soluble}, represents an exactly solvable example for the nonlinear Dirac equation in one-dimensional space \cite{Thirring1958soluble}, and is also used to describe nonlinear optical pulse propagation in Bragg nonlinear optical media \cite{Winful1982optical1,Joseph1989optical2,Aceves1989optical3,Eggleton1996optical4,Eggleton1999optical5}.
The complete integrability of the MT model has presented by means of the inverse scattering transform method \cite{Mikhailov1976integrability,Orfanidis1976soliton}.
Regarding the rogue wave solutions of the MT model, the first-order rogue wave has been constructed through the Darboux method with a matrix version of the Lagrange interpolation method \cite{Degasperis2015Bragg,Degasperis2015darboux}.
Applying this fundamental solution to the coupled mode equations has shown that combining electromagnetically induced transparency with Bragg scattering four-wave mixing may yield rogue waves at low powers.
The higher-order rogue wave solutions have been investigated by using the n-fold Darboux transformation, and the explicit formulas up to third-order with their patterns have been provided in detail \cite{guo2017rogueMT}.
By virtue of the nonrecursive Darboux transformation method, general super rogue wave solutions have been obtained and their structure analyses reveal that rogue waves properties of both components in the coupled MT model is same as that in scalar nonlinear systems except for the spatiotemporal distributions \cite{ye2021super}.
The modulation instability responsible for rogue waves in the MT model has been discussed in \cite{ye2021super}.

In this paper, we derive general rogue wave solutions in the MT model by using the Kadomtsev-Petviashvili (KP) hierarchy reduction method, and these solutions are given explicitly in terms of determinants with the elements given by elementary Schur-polynomial.
In the process of constructing rational solutions, two dimension-reduction conditions and one index-reduction one are proved to be consistent.
It is found that these rational solutions in the MT model depend on the background parameters which do not affect the height of peak but influence the orientations and durations of rogue waves.
Differing from many other coupled systems, there is no additional parameter, thus, the MT model only admits the rogue wave of bright-type.
The $N$-th order rogue wave is shown to be the superposition of $N(N+1)/2$ fundamental ones, and the arrangement pattern depends on the non-reducible parameters.
Particularly, the super rogue wave at each order is achieved simply by setting all internal parameters to zero, and the height of the sole huge peak at order $N$ is $2N+1$ times the background amplitude.
Moreorver, when one of the internal parameters is extreme larger than other ones, rogue wave patterns with certain arrangement shapes are discussed and these patterns are shown to be associated with the root structure of the Yablonskii-Vorob'ev polynomial hierarchy through a linear transformation.

The remainder of the paper is organized as follows. In Section 2, we present general rogue wave solutions to the MT model, which are given in terms of determinants with Schur-polynomial matrix elements.
In Section 3, the rogue wave solutions are derived by using the KP-Toda hierarchy reduction method.
In Section 4, the local structures of fundamental
and higher-order rogue waves are analyzed and illustrated.
Section 5 provides the detailed analyses of rogue wave patterns with one large internal parameter and the connection to the root structure of the Yablonskii-Vorob'ev polynomial hierarchy.
The paper is concluded in Section 6  by a summary and discussion. 

\section{General Rogue Wave Solutions}

The massive Thirring (MT) model is given in terms of the light-cone coordinates \cite{Degasperis2015Bragg,guo2017rogueMT,ye2021super}
\begin{eqnarray}
&& \mathrm{i} u_x + v +\sigma u|v|^2 =0\,, \label{MTa} \\ [5pt]
&& \mathrm{i} v_t + u + \sigma v|u|^2  =0 \label{MTb} \,,
\end{eqnarray}
with $\sigma=\pm1$.
In the optical context, both components $u$ and $v$ represent envelopes of the forward and backward waves respectively \cite{Degasperis2015Bragg}.
On the left-hand sides of equations (\ref{MTa})-(\ref{MTb}), the second and third terms denote the linear coupling and the cross-phase modulation, inducing the dispersion and nonlinearity effects, and their balance leads to the generation of various soliton solutions.

In this work, rogue waves of the MT model (\ref{MTa})-(\ref{MTb}) will be expressed in terms of Schur polynomials.
The elementary Schur polynomials $S_j(\mbox{\boldmath $x$})$ are defined via the generating function
\[ \label{def_Schur}
\sum_{j=0}^{\infty}S_j(\mbox{\boldmath $x$})\lambda^j
=\exp\left(\sum_{j=1}^{\infty}x_j\lambda^j\right),
\]
or more explicitly,
\begin{equation*}
S_0(\mbox{\boldmath $x$})=1, \quad S_1(\mbox{\boldmath $x$})=x_1,
\quad S_2(\mbox{\boldmath $x$})=\frac{1}{2}x_1^2+x_2, \quad \cdots, \quad
S_{j}(\mbox{\boldmath $x$}) =\sum_{l_{1}+2l_{2}+\cdots+ml_{m}=j} \left( \ \prod _{j=1}^{m} \frac{x_{j}^{l_{j}}}{l_{j}!}\right),
\end{equation*}
where $\mbox{\boldmath $x$}=(x_1,x_2,\cdots)$.

The main result of this paper, or the general rogue wave solutions of the MT model (\ref{MTa})-(\ref{MTb}) are given by the following theorem.
\begin{theorem}\label{theorem1}
The MT model (\ref{MTa})-(\ref{MTb}) possesses rogue wave solutions
\begin{equation} \label{var_tran1}
u=\rho_1 \frac{g}{f^{\ast} } e^{\mathrm{i}(1+\sigma \rho_1\rho_2) \left(\frac{\rho_2}{\rho_1}x+\frac{\rho_1}{\rho_2}t\right)}\,, \quad
v=\rho_2 \frac{{h}}{{f}} e^{\mathrm{i}(1+\sigma \rho_1\rho_2) \left(\frac{\rho_2}{\rho_1}x+\frac{\rho_1}{\rho_2}t\right)}\,,
\end{equation}
where
\begin{equation}
f=\sigma_{0,0,0,},\ \ f^*=\sigma_{-1,0,0}, \ \ g= \sigma_{-1,1,0}, \  \  h= \sigma_{-1,0,1},
\end{equation}
and the elements in the determinant $\sigma_{n,k,l} = \det_{1\leq i, j\leq N}\left( m_{2i-1,2j-1}^{(n,k,l)} \right)$ are
\begin{equation}
m^{(n,k,l)}_{i,j}
=\sum^{\min (i,j)}_{\gamma=0}
\frac{1}{4^{\gamma}}
S_{i-\gamma}(\mbox{\boldmath $x$}^+(n,k,l)+\gamma \mbox{\boldmath $s$} )S_{j-\gamma}(\mbox{\boldmath $x$}^-(n,k,l)+\gamma \mbox{\boldmath $s$} ),
\end{equation}
with the vectors $\mbox{\boldmath $x$}^{\pm}(n,k,l)=\Big{(}x_1^{\pm}(n,k,l),x_2^{\pm}(n,k,l),\cdots \Big{)}\equiv \Big{(}x_1^{\pm},x_2^{\pm},\cdots \Big{)}$ being defined by
\begin{eqnarray*}
&&
x^+_r =\alpha_r x+\beta_r t +(n+\frac{1}{2})\theta_r +k \vartheta_r + l \zeta_r +a_r,
\\
&&
x^-_r =\alpha_r x+\beta^*_r t -(n+\frac{1}{2})\theta^*_r -k \vartheta^*_r -l \zeta_r +a^*_r.
\end{eqnarray*}
The parameters $\alpha_r$, $\beta_r$, $\theta_r$, $\vartheta_r$, $\zeta_r$ and $\mbox{\boldmath $s$}=\left(s_1,s_2,\cdots\right)$ are the coefficients from the following expansions
\begin{eqnarray*}
&&
\frac{\rho_2}{\rho_1}[\hat{\rho}(e^\kappa-1)]
=\sum^{\infty}_{r=1}\alpha_r\kappa^r,\
\ \
\frac{\rho_1\rho}{\rho_2}\left[\frac{1}{\hat{\rho}+{\rm i}\rho}-\frac{1}{e^\kappa\hat{\rho}+{\rm i}\rho}\right]
=\sum^{\infty}_{r=1}\beta_r\kappa^r,
\ \
\ln\frac{e^\kappa\hat{\rho}+{\rm i}\rho}{\hat{\rho}+{\rm i}\rho}=\sum^{\infty}_{r=1}\theta_r\kappa^r,
\\
&&
\ln\frac{e^\kappa\hat{\rho}+{\rm i}\sigma\rho_1\rho_2}{\hat{\rho}+{\rm i}\sigma\rho_1\rho_2}
=\sum^{\infty}_{r=1}\vartheta_r\kappa^r,\ \
\ \
\ln e^\kappa =\sum^{\infty}_{r=1}\zeta_r\kappa^r,\ \
\ln\left( \frac{2}{\kappa}\frac{e^\kappa-1}{e^\kappa+1} \right)
=\sum^{\infty}_{r=1} s_r \kappa^r,
\end{eqnarray*}
with $\hat{\rho}=\sqrt{-\sigma\rho_1\rho_2\rho}$ and $\rho=1+\sigma \rho_1\rho_2$.
Here $\rho_1$ and $\rho_2$ are real parameters which satisfy the conditions: $-1<\rho_1\rho_2<0$ for $\sigma=1$ or $0<\rho_1\rho_2<1$ for $\sigma=-1$, and $a_r$ $(r=1,2,\cdots)$ are arbitrary complex parameters.
\end{theorem}

We make three remarks in a row. First, the rogue wave of order $N$ has $N-1$ free irreducible complex parameters, $a_3$, $a_5$, $\cdots$, $a_{2N-1}$, which is the same as that in the NLS equation \cite{Ohta2012NLS}, the DNLS equation \cite{BYang2020DNLS} and the three-wave resonant interaction system \cite{BYang2021TW}.
 The rogue waves of order $N$ contain $2N-1$ complex parameters $a_1$, $a_2$, $\cdots$, $a_{2N-1}$.
However, since the tau function $\sigma_{n,k,l}$ can be rewritten as the summation formula in Ref.\cite{Ohta2012NLS}
which can be treated by the technique in Ref.\cite{BYang2020DNLS},
one finds that the rogue wave is independent of all even-indexed parameters $a_{\mbox{even}}$.
This enables us to take these dummy parameters as zeros, i.e. $a_2=a_4=\cdots=a_{\mbox{even}}=0$.
Of the remaining parameters, $a_1$ can be normalized into zero through a shift of variables $x$ and $t$.
Thus, there only exist $N-1$ free irreducible complex parameters ($a_3$, $a_5$, $\cdots$, $a_{2N-1}$) in the rogue wave of order $N$ .

Second, the rogue wave in the MT model is independent of the parameter $\alpha$ though it is introduced in the process of constructing rogue wave solutions and appears in Theorem \ref{theorem-operator} where solutions are taken in differential operator form. This parameter is finally removed when rogue wave solutions are expressed by elementary Schur polynomials.
This implies that $\alpha$ is a reducible parameter.
In spite of the fact that the DNLS equation and the MT model belong to the same integrable hierarchy,  the parameter $\alpha$ is irreducible in rogue wave solutions of the DNLS equation and it controls the orientation and duration of rogue waves \cite{BYang2020DNLS}. However, for the MT model, as analyzed in Sect.\ref{sect-dynamics}, two free amplitude parameters $\rho_1$ and $\rho_2$ are shown to affect these features.
Differing from other coupled systems in which abundant patterns of rogue wave such as dark and four-petaled flower structures could usually appear, as analyzed in Sect.\ref{sect-fisrtdynamics}, the MT model
only allows the fundamental rogue wave of bright-type which possesses three critical points (one maximum and two minima) and the fixed height of peak.

Third, the rogue wave solutions of the MT model in Theorem \ref{theorem1} are presented in the light cone coordinates,
one can rewrite them in the laboratory coordinates straightforwardly through the independent variables transformations \cite{Degasperis2015darboux,Degasperis2015Bragg,ye2021super}.
The higher-order rogue wave solutions in Ref.\cite{guo2017rogueMT} can be expressed explicitly via the scaling transformations:
\begin{equation}\label{scaling-guo}
u=\pm\frac{1}{\sqrt{-\sigma\mu}} U(\eta,\xi)=\pm\frac{1}{\sqrt{-\sigma\mu}} U(\frac{x}{\mu^2},t),\ \ v=\pm\frac{1}{\mu\sqrt{-\sigma\mu}}V(\eta,\xi)=\pm\frac{1}{\mu\sqrt{-\sigma\mu}}V(\frac{x}{\mu^2},t).
\end{equation}

\begin{prop}\label{theorem-pt}
When $a_r=0$ for all $a\geq1$, the rogue wave solutions in Theorem \ref{theorem1} are parity-time-symmetric, i.e.,
$u^*(-x,-t)=u(x,t)$ and $v^*(-x,-t)=v(x,t)$.
\end{prop}

Indeed, if we set $a_r=0$ in Theorem \ref{theorem1}, the relations $[x^\pm_r(n,k,l)]^*(-x,-t)=-[x^\pm_r(n,k,l)](x,t)$ for all $a\geq1$ hold.
Then applying the same treatment as that in the DNLS equation \cite{BYang2020DNLS}, one can find that $\sigma^*_{n,k,l}(-x,-t)=\sigma_{n,k,l}(x,t)$. Thus, the above Proposition 1
is proved.
From the Proposition 1,
the parity-time-symmetric rogue wave of each order can reach its maximum peak amplitude and
this point is located at the center of such nonlinear wave, i.e., at $(x,t)=(0,0)$.
This special type of rogue wave corresponds to the super rogue wave state in Ref.\cite{ye2021super} which provides the super rogue wave up to second-order.
In addition, the maximum amplitude in such type of rogue wave can be derived by simply setting $x=t=a_r=0$.
Through the direct calculations for five low-order cases, we can conclude that $N$-th order tau functions read
\begin{eqnarray*}
&&
|f(0,0)|_{a_r=0}=\frac{1}{2^{2N^2}}(1+\sigma \rho_1\rho_2)^{N(N+1)/2},
\\
&&
|g(0,0)|_{a_r=0}=|h(0,0)|_{a_r=0}=\frac{2N+1}{2^{2N^2}}(1+\sigma \rho_1\rho_2)^{N(N+1)/2},
\end{eqnarray*}
which leads to the maximum amplitude of the $N$-th order rogue wave
\begin{eqnarray}
|u(0,0)|_{a_r=0}=(2N+1)|\rho_1|,\ \ |v(0,0)|_{a_r=0}=(2N+1)|\rho_2|.
\end{eqnarray}
It implies that the maximum amplitudes in both component are always $2N+1$ times their background.
This property was examined graphically for the first- and second-order rogue waves in \cite{ye2021super}.
Hence, the maximum amplitude of the super rogue wave in the MT model are the same as that in scaling integrable systems such as the NLS and DNLS equations.
On the other hand, as reported in previous studies regarding the super rogue waves in coupled integrable systems \cite{guo2011roguecpl,baronio2012solutions,baronio2014vector,chen2014dark,chow2013rogue,chen2018jpsj,chen2019pre,Degasperis2013rational,wang2015higher,chen2013watch,zhang2018three,BYang2021TW}, the additional degrees of freedom could give rise to the varying maximum peak amplitude.

\section{Derivation of rogue wave solutions}
In this section, general rogue wave solutions will be derived by using the KP hierarchy reduction method.
Prior to the tedious derivation, we list the main steps as shown in the subsequent subsections.
First, we introduce the dependent variable transformations
\begin{equation} \label{var_tran2}
u=\rho_1 \frac{g}{f^{\ast} } e^{\mathrm{i}(1+\sigma \rho_1\rho_2) \left(\frac{\rho_2}{\rho_1}x+\frac{\rho_1}{\rho_2}t\right)}\,, \quad
v=\rho_2 \frac{{h}}{{f}} e^{\mathrm{i}(1+\sigma \rho_1\rho_2) \left(\frac{\rho_2}{\rho_1}x+\frac{\rho_1}{\rho_2}t\right)}\,,
\end{equation}
where $f,g,h$ are complex functions, and $\rho_1,\rho_2$ are real constants.
The MT model (\ref{MTa})--(\ref{MTb}) is transformed into the following bilinear equations
\begin{eqnarray}
&&  (\mathrm{i} D_{x}- \frac{\rho_2}{\rho_1}) g \cdot f= - \frac{\rho_2}{\rho_1}  h f^{\ast}  \,,  \label{MTdkBL1} \\
&&  (\mathrm{i} D_{x}  - \sigma \rho^2_2 ) f \cdot f^{\ast } =-\sigma \rho^2_2 hh^{\ast} \,, \label{MTdkBL2}  \\
&&  (\mathrm{i} D_{t} -\frac{\rho_1}{\rho_2}) h \cdot f^{\ast}  = - \frac{\rho_1}{\rho_2} g f  \,,  \label{MTdkBL3} \\
&&  (\mathrm{i} D_{t}  - \sigma \rho^2_1 ) f^{\ast} \cdot f  =-\sigma \rho^2_1 gg^{\ast} \,, \label{MTdkBL4}
\end{eqnarray}
where $D$ is the Hirota's bilinear differential operator defined by
\begin{equation*}
D_s^n D_y^m f\cdot g=\left(\frac{\partial}{\partial s} -\frac{\partial}{%
\partial s^{\prime }}\right)^n \left(\frac{\partial}{\partial y} -\frac{%
\partial}{\partial y^{\prime }}\right)^m f(y,s)g(y^{\prime },s^{\prime
})|_{y=y^{\prime }, s=s^{\prime }}\,.
\end{equation*}

Next, we consider a set of higher-dimensional bilinear equations in extended KP hierarchy which includes negative flows
\begin{eqnarray}
&&
(D_{x_1}+a) \tau_{n,k+1,l}\cdot \tau_{n+1,k,l} =a \tau_{n+1,k+1,l} \tau_{n,k,l}, \label{KPbilinear1} \\
&&
(bD_{x_{-1}}+1) \tau_{n,k,l+1}\cdot \tau_{n,k,l} = \tau_{n-1,k,l+1} \tau_{n+1,k,l}, \label{KPbilinear2} \\
&&
(aD_{t_{a}}-1)\tau_{n+1,k,l}\cdot \tau_{n,k,l}=-\tau_{n+1,k-1,l}\tau_{n,k+1,l}, \label{KPbilinear3} \\
&&
(bD_{t_{b}}-1)\tau_{n+1,k,l}\cdot \tau_{n,k,l}=-\tau_{n+1,k,l-1}\tau _{n,k,l+1}\,, \label{KPbilinear4}
\end{eqnarray}
which allow a wide class of algebraic solutions in terms of Gram determinant.
Based on these algebraic solutions, we restrict them to satisfy the dimension- and index-reduction conditions:
\begin{eqnarray}
\label{dim-red-con1}&& [\partial_{x_1}-b(a-b)\partial_{t_b}]\tau_{ n,k,l }=C_1\tau_{ n,k,l },   \\
\label{dim-red-con2}&& [\partial_{x_{-1}}+\frac{a-b}{b}\partial_{t_a}]\tau_{ n,k,l }=C_2\tau_{ n,k,l },\\
\label{index-red-con3}&& \tau_{n+1,k+1,l-1 }=e^{{\rm i}\kappa_0} \tau_{ n,k,l },
\end{eqnarray}
where $C_1$, $C_2$ and $\kappa_0$ are real constants.
Then such algebraic solutions satisfy the (1+1)-dimensional bilinear equations:
\begin{eqnarray}
&& (D_{x_1}+a) \tau_{n,k+1,l}\cdot \tau_{n+1,k,l} =a e^{{\rm i}\kappa} \tau_{n,k,l+1} \tau_{n,k,l}, \label{dark-before-eeq1}\\
&& (bD_{x_{-1}}+1) \tau_{n,k,l+1}\cdot \tau_{n,k,l} = e^{-{\rm i}\kappa} \tau_{n,k+1,l} \tau_{n+1,k,l},  \label{dark-before-eeq3}\\
&& \left[-\frac{ab}{(a-b)}D_{x_{-1}}-1\right]\tau_{n+1,k,l}\cdot \tau_{n,k,l}=-\tau_{n+1,k-1,l}\tau_{n,k+1,l},\\
&& \left[\frac{1}{a-b}D_{x_{1}}-1\right]\tau_{n+1,k,l}\cdot \tau_{n,k,l}=-\tau_{n+1,k,l-1}\tau _{n,k,l+1}\,. \label{dark-before-eeq4}
\end{eqnarray}

Finally, by setting the coordinate transformations
\begin{equation}
x_1=-\frac{\rho_2}{{\rm i}a\rho_1}x,\ \ x_{-1}=-\frac{\rho_1 b}{{\rm i}\rho_2}t,\ \
\end{equation}
i.e.,
\begin{equation*}
\partial_{x_1}= -\frac{{\rm i}a\rho_1}{\rho_2}  \partial_{x},\ \ \partial_{x_{-1}}=-\frac{{\rm i}\rho_2}{\rho_1 b} \partial_{t},
\end{equation*}
with the relation $b=a(1+\sigma \rho_1\rho_2)$, and taking the variable transformations $(n=-1,k=0,l=0)$
\begin{equation*}
\tau_{0,0,0,}=f, \ \tau_{-1,0,0}=f^*, \  \tau_{-1,1,0}=e^{{\rm i}\kappa_0}g, \  \tau_{0,-1,0}=e^{-{\rm i}\kappa_0}g^*, \  \tau_{-1,0,1}=h, \  \tau_{0,0,-1}=h^*,
\end{equation*}%
we arrive at exactly the bilinear equations (\ref{MTdkBL1})--(\ref{MTdkBL4}).

\subsection{Gram determinant solution for a higher-dimensional bilinear system }

In this subsection, we present Gram determinant solution for the higher-dimensional bilinear equations (\ref{KPbilinear1})-(\ref{KPbilinear4}).

\begin{lem}\label{lem1}
Let $m^{(n,k,l)}_{ij}$, $\varphi^{(n,k,l)}_{i}$ and $\psi^{(n,k,l)}_{j}$ be functions of variables $x_1$, $x_{-1}$, $t_a$ and $t_b$
satisfying the differential and difference relations as follows,
\begin{eqnarray}\label{rules1}
\nonumber &&
\partial_{x_1}m^{(n,k,l)}_{ij}=\mu_0\varphi^{(n+1,k,l)}_{i}\psi^{(n,k,l)}_{j},
\\
\nonumber &&
\partial_{x_{-1}}m^{(n,k,l)}_{ij}=-\mu_0\varphi^{(n,k,l)}_{i}\psi^{(n+1,k,l)}_{j},
\\
\nonumber &&
\partial_{t_a}m^{(n,k,l)}_{ij}=-\mu_0\varphi^{(n+1,k-1,l)}_{i}\psi^{(n,k+1,l)}_{j},
\\
&&
\partial_{t_b}m^{(n,k,l)}_{ij}=-\mu_0\varphi^{(n+1,k,l-1)}_{i}\psi^{(n,k,l+1)}_{j},
\\
\nonumber &&
m^{(n+1,k,l)}_{ij}=m^{(n,k,l)}_{ij}+\mu_0\varphi^{(n+1,k,l)}_{i}\psi^{(n+1,k,l)}_{j},
\\
\nonumber &&
m^{(n,k+1,l)}_{ij}=m^{(n,k,l)}_{ij}+\mu_0\varphi^{(n+1,k,l)}_{i}\psi^{(n,k+1,l)}_{j},
\\
\nonumber &&
m^{(n,k,l+1)}_{ij}=m^{(n,k,l)}_{ij}+\mu_0\varphi^{(n+1,k,l)}_{i}\psi^{(n,k,l+1)}_{j},
\end{eqnarray}
and
\begin{eqnarray}\label{rules2}
\nonumber &&
\partial_{x_1}\varphi^{(n,k,l)}_{i}=\varphi^{(n+1,k,l)}_{i},\ \ \partial_{x_1}\psi^{(n,k,l)}_{j}=-\psi^{(n-1,k,l)}_{j},
\\
\nonumber &&
\partial_{x_{-1}}\varphi^{(n,k,l)}_{i}=\varphi^{(n-1,k,l)}_{i},\ \ \partial_{x_{-1}}\psi^{(n,k,l)}_{j}=-\psi^{(n+1,k,l)}_{j},
\\
 &&
\partial_{t_a}\varphi^{(n,k,l)}_{i}=\varphi^{(n,k-1,l)}_{i},\ \ \partial_{t_a}\psi^{(n,k,l)}_{j}=-\psi^{(n,k+1,l)}_{j},
\\
\nonumber &&
\partial_{t_b}\varphi^{(n,k,l)}_{i}=\varphi^{(n,k,l-1)}_{i},\ \ \partial_{t_b}\psi^{(n,k,l)}_{j}=-\psi^{(n,k,l+1)}_{j},
\\
\nonumber &&
\varphi^{(n+1,k,l)}_{i}=\varphi^{(n,k+1,l)}_{i}+a\varphi^{(n,k,l)}_{i},\ \ \psi^{(n+1,k,l)}_{j}=\psi^{(n,k+1,l)}_{i}-a\psi^{(n+1,k+1,l)}_{j},
\\
\nonumber &&
\varphi^{(n+1,k,l)}_{i}=\varphi^{(n,k,l+1)}_{i}+b\varphi^{(n,k,l)}_{i},\ \ \psi^{(n+1,k,l)}_{j}=\psi^{(n,k,l+1)}_{i}-b\psi^{(n+1,k,l+1)}_{j}.
\end{eqnarray}

Then it is verified that the determinant
\begin{equation}
\tau_{ n,k,l }= \det_{1\leq i,j\leq N}\left(m^{(n,k,l)}_{ij} \right),
\end{equation}
satisfy the following  bilinear equations in extended KP hierarchy
\begin{eqnarray}
&&
(D_{x_1}+a) \tau_{n,k+1,l}\cdot \tau_{n+1,k,l} =a \tau_{n+1,k+1,l} \tau_{n,k,l}, \label{KKPbilinear1} \\
&&
(bD_{x_{-1}}+1) \tau_{n,k,l+1}\cdot \tau_{n,k,l} = \tau_{n-1,k,l+1} \tau_{n+1,k,l}, \label{KKPbilinear2} \\
&&
(aD_{t_{a}}-1)\tau_{n+1,k,l}\cdot \tau_{n,k,l}=-\tau_{n+1,k-1,l}\tau_{n,k+1,l}, \label{KKPbilinear3} \\
&&
(bD_{t_{b}}-1)\tau_{n+1,k,l}\cdot \tau_{n,k,l}=-\tau_{n+1,k,l-1}\tau _{n,k,l+1}\,. \label{KKPbilinear4}
\end{eqnarray}
\end{lem}

\emph{Proof.} By utilizing rules (\ref{rules1}) and (\ref{rules2}), one can check that the derivatives
and shifts of the tau function are expressed by the bordered determinants as follows:
\begin{eqnarray*}
&&
\tau_{n+1,k,l}=\left\vert
\begin{array}{cc}
m^{(n,k,l)}_{ij} & \varphi^{(n+1,k,l)}_{i} \\
-\mu_0\psi^{(n+1,k,l)}_{j} &  1
\end{array}\right\vert,
\ \
\tau_{n,k+1,l}=\left\vert
\begin{array}{cc}
m^{(n,k,l)}_{ij} & \varphi^{(n+1,k,l)}_{i} \\
-\mu_0\psi^{(n,k+1,l)}_{j} &  1
\end{array}\right\vert,
\\
&&
\tau_{n,k,l+1}=\left\vert
\begin{array}{cc}
m^{(n,k,l)}_{ij} & \varphi^{(n+1,k,l)}_{i} \\
-\mu_0\psi^{(n,k,l+1)}_{j} &  1
\end{array}\right\vert,\ \
\partial_{x_{-1}}\tau_{n,k,l}=\left\vert
\begin{array}{cc}
m^{(n,k,l)}_{ij} & \varphi^{(n,k,l)}_{i} \\
\mu_0\psi^{(n+1,k,l)}_{j} &  0
\end{array}\right\vert,
\\
&&
\partial_{t_a}\tau_{n,k,l}=\left\vert
\begin{array}{cc}
m^{(n,k,l)}_{ij} & \varphi^{(n+1,k-1,l)}_{i} \\
\mu_0\psi^{(n,k+1,l)}_{j} &  0
\end{array}\right\vert,
\ \
\partial_{t_b}\tau_{n,k,l}=\left\vert
\begin{array}{cc}
m^{(n,k,l)}_{ij} & \varphi^{(n+1,k,l-1)}_{i} \\
\mu_0\psi^{(n,k,l+1)}_{j} &  0
\end{array}\right\vert,
\\
&&
\partial_{x_1}\tau_{n,k+1,l}=\left\vert
\begin{array}{cc}
m^{(n,k,l)}_{ij} & \varphi^{(n+1,k+1,l)}_{i} \\
-\mu_0\psi^{(n,k+1,l)}_{j} &  0
\end{array}\right\vert,
\ \
\partial_{x_1}\tau_{n+1,k,l}=\left\vert
\begin{array}{cc}
m^{(n,k,l)}_{ij} & \varphi^{(n+2,k,l)}_{i} \\
-\mu_0\psi^{(n+1,k,l)}_{j} &  0
\end{array}\right\vert,
\\
&&
\tau_{n+1,k+1,l}=\left\vert
\begin{array}{ccc}
m^{(n,k,l)}_{ij} & \varphi^{(n+1,k+1,l)}_{i} & \varphi^{(n+1,k,l)}_{i}  \\
-\mu_0\psi^{(n+1,k+1,l)}_{j} &  1 & 0 \\
-\mu_0\psi^{(n,k+1,l)}_{j} &  0 & 1
\end{array}\right\vert,
\ \
\tau_{n-1,k,l+1}=\left\vert
\begin{array}{cc}
m^{(n,k,l)}_{ij} & -b\varphi^{(n,k,l)}_{i} \\
\mu_0\psi^{(n,k,l+1)}_{j} &  1
\end{array}\right\vert,
\\
&&
\tau_{n+1,k-1,l}=\left\vert
\begin{array}{cc}
m^{(n,k,l)}_{ij} & \varphi^{(n+1,k-1,l)}_{i} \\
\mu_0a\psi^{(n+1,k,l)}_{j} &  1
\end{array}\right\vert,
\ \
\tau_{n+1,k,l-1}=\left\vert
\begin{array}{cc}
m^{(n,k,l)}_{ij} & \varphi^{(n+1,k,l-1)}_{i} \\
\mu_0b\psi^{(n+1,k,l)}_{j} &  1
\end{array}\right\vert,
\\
&&
\partial_{x_{-1}}\tau_{n,k,l+1}
=\left\vert
\begin{array}{ccc}
m^{(n,k,l)}_{ij} & \varphi^{(n+1,k,l)}_{i} & \varphi^{(n,k,l)}_{i} \\
-\mu_0\psi^{(n,k,l+1)}_{j} &  1 & 0 \\
\mu_0\psi^{(n+1,k,l)}_{j} &  0 & 0
\end{array}\right\vert
+
\left\vert
\begin{array}{cc}
m^{(n,k,l)}_{ij} & \varphi^{(n,k,l)}_{i} \\
-\mu_0\psi^{(n,k,l+1)}_{j} &  0
\end{array}\right\vert
+
\left\vert
\begin{array}{cc}
m^{(n,k,l)}_{ij} & \varphi^{(n+1,k,l)}_{i} \\
\mu_0\psi^{(n+1,k,l+1)}_{j} &  0
\end{array}\right\vert,
\\
&&
\partial_{t_a}\tau_{n+1,k,l}
=\left\vert
\begin{array}{ccc}
m^{(n,k,l)}_{ij} & \varphi^{(n+1,k,l)}_{i} & \varphi^{(n+1,k-1,l)}_{i} \\
-\mu_0\psi^{(n+1,k,l)}_{j} &  1 & 0 \\
\mu_0\psi^{(n,k+1,l)}_{j} &  0 & 0
\end{array}\right\vert
-
\left\vert
\begin{array}{cc}
m^{(n,k,l)}_{ij} & \varphi^{(n+1,k-1,l)}_{i} \\
\mu_0\psi^{(n+1,k,l)}_{j} &  0
\end{array}\right\vert
+
\left\vert
\begin{array}{cc}
m^{(n,k,l)}_{ij} & \varphi^{(n+1,k,l)}_{i} \\
\mu_0\psi^{(n+1,k+1,l)}_{j} &  0
\end{array}\right\vert,
\\
&&
\partial_{t_b}\tau_{n+1,k,l}
=\left\vert
\begin{array}{ccc}
m^{(n,k,l)}_{ij} & \varphi^{(n+1,k,l)}_{i} & \varphi^{(n+1,k,l-1)}_{i} \\
-\mu_0\psi^{(n+1,k,l)}_{j} &  1 & 0 \\
\mu_0\psi^{(n,k,l+1)}_{j} &  0 & 0
\end{array}\right\vert
-
\left\vert
\begin{array}{cc}
m^{(n,k,l)}_{ij} & \varphi^{(n+1,k,l-1)}_{i} \\
\mu_0\psi^{(n+1,k,l)}_{j} &  0
\end{array}\right\vert
+
\left\vert
\begin{array}{cc}
m^{(n,k,l)}_{ij} & \varphi^{(n+1,k,l)}_{i} \\
\mu_0\psi^{(n+1,k,l+1)}_{j} &  0
\end{array}\right\vert.
\end{eqnarray*}
Applying the Jacobi identity of determinants to these bordered determinants, the four bilinear equations (\ref{KKPbilinear1})-(\ref{KKPbilinear4}) are satisfied.

In order to construct the algebraic solutions, we first introduce $m^{(n,k,l)}$, $\varphi^{(n,k,l)}$ and $\psi^{(n,k,l)}$ as
\begin{eqnarray}
&&
m^{(n,k,l)}=\frac{\mathrm{i}p }{p +q }
\left( -\frac{p}{q}\right) ^{n}
\left( -\frac{p-a}{q+a}\right) ^{k}
\left( -\frac{p-b}{q+b}\right) ^{l}
e^{\xi+\eta},
\\
&&
\varphi^{(n,k,l)}= p^n (p-a)^k (p-b)^l e^{\xi},
\\
&&
\psi^{(n,k,l)}=(-q)^{-n} [-(q+a)]^{-k} [-(q+b)]^{-l} e^{\eta},
\end{eqnarray}
and
\begin{eqnarray*}
&&
\xi=\frac{1}{p}x_{-1}+p x_{1}+\frac{1}{p -a}t_{a}+\frac{1}{p -b}t_{b}+\xi_0,
\\
&&
\eta=\frac{1}{q}x_{-1}+q x_{1}+\frac{1}{q+a}t_{a}+\frac{1}{q+b}t_{b}+\eta_0,
\end{eqnarray*}
where $\mu_0={\rm i}$, and $p,q,\xi_0,\eta_0,a,b$ are complex constants.
It is easy to find that these functions satisfy differential and difference rules (\ref{rules1}) and (\ref{rules2}) without indices $i$ and $j$.

Then, we define the elements
\begin{equation}
m_{ij}^{(n,k,l)}=\mathcal{A}_i \mathcal{B}_{j} m^{(n,k,l)}, \quad
\varphi_i^{(n,k,l)}=\mathcal{A}_i\varphi^{(n,k,l)}, \quad
\psi_j^{(n,k,l)}=\mathcal{B}_{j}\psi^{(n,k,l)},
\end{equation}
where $\mathcal{A}_{i}$ and $\mathcal{B}_{j}$ are differential operators with respect to $p$ and $q$ respectively as
\begin{eqnarray}
\mathcal{A}_{i}=\frac{1}{ i !}\left[f_1(p)\partial_{p}\right]^{i}, \quad
\mathcal{B}_{j}=\frac{1}{ j !}\left[f_2(q)\partial_{q}\right]^{j},
\end{eqnarray}
and $f_1(p)$, $f_2(p)$ are arbitrary functions which will be determined by the dimensional reduction \cite{BYang2020DNLS,BYang2021TW} in the subsequent  section \ref{dimension-reduction}. 
Since operators $\mathcal{A}_{i}$ and $\mathcal{B}_{j}$ commute with operators $\partial_{x_1}$, $\partial_{x_{-1}}$, $\partial_{t_a}$
and $\partial_{t_b}$, these functions $m_{ij}^{(n,k,l)}$, $\varphi_i^{(n,k,l)}$ and $\psi_j^{(n,k,l)}$ still obey the differential and difference  rules (\ref{rules1}) and (\ref{rules2}).
From Lemma \ref{lem1}, it is known that for an arbitrary sequence of indices $(i_1,i_2,\cdots,i_N; j_1,j_2,\cdots,j_N)$, the determinant
\begin{equation}\label{general-det}
\tau_{n,k,l}=\det_{1\le\nu,\mu\le N}\left( m_{i_\nu,j_\mu}^{(n,k,l)}\right),
\end{equation}
satisfies the higher-dimensional bilinear system (\ref{KKPbilinear1})-(\ref{KKPbilinear4}).

\subsection{Dimensional reduction}
\label{dimension-reduction}

According to the generalized dimensional reduction technique developed in \cite{BYang2021TW},
we introduce the following linear differential operators
\begin{equation}
\mathcal{L}_1=\partial_{x_1}-b(a-b)\partial_{t_b},\ \
\mathcal{L}_2=\partial_{x_{-1}}+\frac{a-b}{b}\partial_{t_a},
\end{equation}
by which the dimensional reduction conditions (\ref{dim-red-con1}) and (\ref{dim-red-con2}) become
\begin{equation}\label{dim-reduction1}
\mathcal{L}_1\tau_{ n,k,l }=C_1\tau_{ n,k,l },\ \ \mathcal{L}_2\tau_{ n,k,l }=C_2\tau_{ n,k,l }.
\end{equation}
It is straightforward  that
\begin{eqnarray}
&&
\mathcal{L}_1m_{ij}^{(n,k,l)}= \mathcal{A}_i \mathcal{B}_{j} \mathcal{L}_1 m^{(n,k,l)}
= \mathcal{A}_i \mathcal{B}_{j} \left [\mathcal{Q}_{11}(p) + \mathcal{Q}_{12}(q) \right]m^{(n,k,l)},
\\
&&
\mathcal{L}_2m_{ij}^{(n,k,l)}= \mathcal{A}_i \mathcal{B}_{j} \mathcal{L}_2 m^{(n,k,l)}
= \mathcal{A}_i \mathcal{B}_{j} \left [\mathcal{Q}_{21}(p) + \mathcal{Q}_{22}(q) \right]m^{(n,k,l)},
\end{eqnarray}
where
\begin{eqnarray*}
&&
\mathcal{Q}_{11}(p)=p-b+\frac{b(b-a)}{p-b},\ \
\mathcal{Q}_{12}(q)=q+b+\frac{b(b-a)}{q+b},
\\
&&
\mathcal{Q}_{21}(p)=\frac{1}{p}+\frac{a-b}{b(p-a)},\ \
\mathcal{Q}_{22}(q)=\frac{1}{q}+\frac{a-b}{b(q+a)}.
\end{eqnarray*}

Using the Leibnitz rule, one has the operator relations \cite{BYang2021TW}
\begin{equation}\label{Leibnitz-rule}
\mathcal{A}_i \mathcal{Q}_{s,1}(p)=\sum^i_{\mu=0}\frac{1}{\mu !}\left[\left(f_1\partial_p\right)^\mu \mathcal{Q}_{s,1}(p) \right] \mathcal{A}_{i-\mu},\ \
\mathcal{B}_i \mathcal{Q}_{s,2}(q)=\sum^i_{\mu=0}\frac{1}{\mu !}\left[\left(f_2\partial_q\right)^\mu \mathcal{Q}_{s,2}(q) \right] \mathcal{B}_{i-\mu},\ \ (s=1,2),
\end{equation}
which suggest
\begin{eqnarray}\label{sum1}
\mathcal{L}_s m_{ij}^{(n,k,l)}=\sum^i_{\mu=0}\frac{1}{\mu !}\left[\left(f_1\partial_p\right)^\mu \mathcal{Q}_{s,1}(p) \right] m^{(n,k,l)}_{i-\mu,j}
+\sum^j_{\nu=0}\frac{1}{\nu !}\left[\left(f_2\partial_q\right)^\nu \mathcal{Q}_{s,2}(q) \right] m^{(n,k,l)}_{i,j-\nu},\ \
\ \ (s=1,2).
\end{eqnarray}

Next, the specific functions $[f_1(p), f_2(q)]$ and values of $(p,q)$ need to be determined to guarantee that
coefficients of certain indices vanish in the above summation.
To this end, we solve the first two algebraic equations
\begin{equation}\label{Q1-eq1}
\mathcal{Q}'_{11}(p)=0,\ \ \mathcal{Q}'_{12}(q)=0,
\end{equation}
and get the following simple roots:
\begin{equation}\label{roots1}
p_0=\sqrt{b(b-a)}+b,\ \ q_0=\sqrt{b(b-a)}-b.
\end{equation}
It is noted that $p_0$ and $q_0$ are also simple roots of equations $\mathcal{Q}'_{21}(p)=0$ and $\mathcal{Q}'_{22}(q)=0$ respectively, since $\mathcal{Q}_{21}(p)$ and $\mathcal{Q}_{22}(q)$ are associated with $\mathcal{Q}_{11}(p)$ and $\mathcal{Q}_{12}(q)$ through the simple relations
\begin{equation}\label{Q2link}
\mathcal{Q}_{21}(p)=\frac{a}{b}[\mathcal{Q}_{11}(p)-a+2b]^{-1}, \ \ \mathcal{Q}_{22}(q)=\frac{a}{b}[\mathcal{Q}_{12}(q)+a-2b]^{-1}.
\end{equation}
Hence, the terms with $\mu=\nu=1$ on the r.h.s of (\ref{sum1}) vanish at the point $(p_0,q_0)$.

Following the steps in the simple root's case in Ref.\cite{BYang2021TW}, we need to solve the differential equations
\begin{equation}\label{f1f2-equ}
\left(f_1\partial_p\right)^2 \mathcal{Q}_{11}(p)= \mathcal{Q}_{11}(p),\ \
\left(f_2\partial_q\right)^2 \mathcal{Q}_{12}(q)= \mathcal{Q}_{12}(q).
\end{equation}
Then the following functions can be derived
\begin{equation}\label{W12-exp}
\mathcal{W}_1(p)=\left(\frac{p-b}{\sqrt{b(b-a)}}\right)^{\pm 1},\ \ \mathcal{W}_2(q)=\left(\frac{q+b}{\sqrt{b(b-a)}}\right)^{\pm 1}.
\end{equation}
and
\begin{equation}\label{f12-exp}
f_1(p)=\frac{\mathcal{W}_1(p)}{\mathcal{W}'_1(p)}=\pm(p-b),\ \ f_2(q)=\frac{\mathcal{W}_2(p)}{\mathcal{W}'_2(p)}=\pm(q+b).
\end{equation}
Because the above two signs yield equivalent rogue wave solutions, we choose positive sign in the following derivation.
From the conditions (\ref{Q1-eq1}) and (\ref{f1f2-equ}), we can find that
\begin{eqnarray}
\label{contiguity1}&&
\left.\mathcal{L}_1 m_{ij}^{(n,k,l)}\right|_{p=p_{0},  q=q_{0}}=\mathcal{Q}_{11}(p_0) \sum^i_{\begin{subarray}{c} \mu=0,\\ \mu: even \end{subarray}}\frac{1}{\mu !} \left.m^{(n,k,l)}_{i-\mu,j}\right|_{p=p_{0},  q=q_{0}}
+\mathcal{Q}_{12}(q_0)\sum^j_{\begin{subarray}{c} \nu=0,\\ \nu: even \end{subarray}}\frac{1}{\nu !} \left.m^{(n,k,l)}_{i,j-\nu}\right|_{p=p_{0},  q=q_{0}},
\end{eqnarray}

In order to prove the second dimensional reduction, noticing that $\mathcal{Q}_{21}(p)$ can be expressed as a function of $\mathcal{Q}_{11}(p)$, we consider a general function
$F[\mathcal{Q}_{11}(p)]\equiv F(\mathcal{Q}_{11})$. 
By using the Fa\`{a} di Bruno formula and the relation $f_1\partial_p=\partial_{\ln\mathcal{W}_1} $,
 we obtain
\begin{eqnarray}\label{high-der}
\left(f_1\partial_p\right)^lF(\mathcal{Q}_{11})= \partial^l_{\ln\mathcal{W}_1}F(\mathcal{Q}_{11})
=\sum_{m_1+2m_2+\cdots+lm_l=l}  \frac{ \frac{d^{\hat{m}}F(\mathcal{Q}_{11})}{d \mathcal{Q}^{\hat{m}}_{11}} \prod^l_{j=1}\left[ (f_1\partial_p)^j \mathcal{Q}_{11} \right]^{m_j}}{(l!)^{-1} \prod^l_i m_i!(i!)^{m_i}}
.
\end{eqnarray}
where $\hat{m}=\sum^l_{i=1} m_i$.
Furthermore, by using the conditions (\ref{Q1-eq1}) and (\ref{f1f2-equ}), one finds that
\begin{eqnarray}
\label{Ff-1}&&
\left(f_1\partial_p\right)^{l}F[\mathcal{Q}_{11}(p_0)]=0,\ \ ( l \mbox{ is odd}),
\\
\label{Ff-2}&&
\left(f_1\partial_p\right)^{l}F[\mathcal{Q}_{11}(p_0)]=
\left.\sum_{\begin{subarray}{c} 2m_2+\cdots+lm_l=l,\\ m_1=m_3=\cdots=0 \end{subarray}}
 \frac{ \frac{d^{\hat{m}}F(\mathcal{Q}_{11})}{d \mathcal{Q}^{\hat{m}}_{11}} \prod^l_{j=1}\left[ (f_1\partial_p)^j \mathcal{Q}_{11} \right]^{m_j}}{(l!)^{-1} \prod^l_i m_i!(i!)^{m_i}}
\right|_{p=p_0} \triangleq \mathcal{C}_{1,l}\left(F[\mathcal{Q}_{11}(p_0)]\right),\ \ ( l \mbox{ is even}).
\end{eqnarray}
The similar calculation for a general function $F[\mathcal{Q}_{12}(q)]\equiv F(\mathcal{Q}_{12})$ gives
\begin{eqnarray}
\label{Ff-3}&&
\left(f_2\partial_q\right)^{l}F[\mathcal{Q}_{12}(q_0)]=0,\ \ ( l \mbox{ is odd}),
\\
\label{Ff-4}&&
\left(f_2\partial_q\right)^{l}F[\mathcal{Q}_{12}(q_0)]=
\left.\sum_{\begin{subarray}{c} 2m_2+\cdots+lm_l=l,\\ m_1=m_3=\cdots=0 \end{subarray}}
 \frac{ \frac{d^{\hat{m}}F(\mathcal{Q}_{12})}{d \mathcal{Q}^{\hat{m}}_{12}} \prod^l_{j=1}\left[ (f_2\partial_q)^j \mathcal{Q}_{12} \right]^{m_j}}{(l!)^{-1} \prod^l_i m_i!(i!)^{m_i}}
\right|_{p=p_0} \triangleq \mathcal{C}_{2,l}\left(F[\mathcal{Q}_{12}(q_0)]\right),\ \ ( l \mbox{ is even}).
\end{eqnarray}

Applying the above formulas (\ref{Ff-1})-(\ref{Ff-4}) to the specific functions $F[\mathcal{Q}_{11}(p)]=\mathcal{Q}_{21}(p)$ and $F[\mathcal{Q}_{12}(q)]=\mathcal{Q}_{22}(q)$, it follows that
\begin{eqnarray}
\label{contiguity2}&&
\left.\mathcal{L}_2 m_{ij}^{(n,k,l)}\right|_{p=p_{0},  q=q_{0}}
=\sum^i_{\begin{subarray}{c} \mu=0,\\ \mu: even \end{subarray}}\frac{\mathcal{C}_{1,\mu}[\mathcal{Q}_{21}(p_0)] }{\mu !} \left.m^{(n,k,l)}_{i-\mu,j}\right|_{p=p_{0},  q=q_{0}}
+\sum^j_{\begin{subarray}{c} \nu=0,\\ \nu: even \end{subarray}}\frac{\mathcal{C}_{2,\nu}[\mathcal{Q}_{22}(q_0)] }{\nu !} \left.m^{(n,k,l)}_{i,j-\nu}\right|_{p=p_{0},  q=q_{0}}.
\end{eqnarray}
On the r.h.s. of above equation, the coefficients in the first term of two summations are $\mathcal{C}_{1,0}[\mathcal{Q}_{21}(p_0)]=\mathcal{Q}_{21}(p_0)$ and $\mathcal{C}_{2,0}[\mathcal{Q}_{22}(q_0)]=\mathcal{Q}_{22}(q_0)$, respectively.

Next, we restrict the general determinant (\ref{general-det}) to
\begin{equation}\label{general-det1}
\tau_{n,k,l} = \det_{1\leq i, j\leq N}\left(\left. m_{2i-1,2j-1}^{(n,k,l)} \right|_{p=p_{0}, q=q_{0}} \right).
\end{equation}
By using the contiguity relations (\ref{contiguity1}) and (\ref{contiguity2}) as in Ref.\cite{Ohta2012NLS}, we obtain
\begin{eqnarray}
&&
\mathcal{L}_1\tau_{n,k,l}=\left[\mathcal{Q}_{11}(p_0)+ \mathcal{Q}_{12}(q_0) \right] N \tau_{n,k,l}=4\sqrt{b(b-a)}N \tau_{n,k,l},
\\
&&
\mathcal{L}_2\tau_{n,k,l}=\left[\mathcal{Q}_{21}(p_0)+ \mathcal{Q}_{22}(q_0) \right] N \tau_{n,k,l}=-\frac{4\sqrt{b(b-a)}}{ab}N \tau_{n,k,l},
\end{eqnarray}
which imply that the tau function (\ref{general-det1}) satisfies the dimensional reduction conditions (\ref{dim-reduction1}).

\subsection{Index reduction}

Similar to the dimensional reduction procedure in the above subsection, we introduce the difference operator
\begin{equation}
\Delta \tau_{n,k,l} =\tau_{n+1,k+1,l-1},
\end{equation}
so that the index reduction condition (\ref{index-red-con3}) becomes
\begin{equation}\label{index-reduction1}
\Delta \tau_{n,k,l}=e^{{\rm i}\kappa_0} \tau_{n,k,l}.
\end{equation}
From the definition of element $m_{ij}^{(n,k,l)}$, we can find
\begin{equation}
\Delta m_{ij}^{(n,k,l)}= \mathcal{A}_i \mathcal{B}_{j} \Delta m^{(n,k,l)}
= \mathcal{A}_i \mathcal{B}_{j} \left [\mathcal{Q}_{31}(p) \times \mathcal{Q}_{32}(q) \right]m^{(n,k,l)},
\end{equation}
where
\begin{equation}\label{Q3link}
\mathcal{Q}_{31}(p)= \frac{p(p-a)}{p-b}=\mathcal{Q}_{11}(p)-a+2b,\ \ \mathcal{Q}_{32}(q)=-\frac{q+b}{q(q+a)}=-[\mathcal{Q}_{12}(q)+a-2b]^{-1},
\end{equation}
which yield
\begin{equation}
\left.\mathcal{Q}'_{31}(p)\right|_{p=p_{0},  q=q_{0}}=0,\ \
\left.\mathcal{Q}'_{32}(q)\right|_{p=p_{0},  q=q_{0}}=0.
\end{equation}

Furthermore, applying above formulas (\ref{Ff-1})-(\ref{Ff-4}) to the specific functions $F[\mathcal{Q}_{11}(p)]=\mathcal{Q}_{31}(p)$ and $F[\mathcal{Q}_{12}(q)]=\mathcal{Q}_{32}(q)$ and utilizing general operator relations (\ref{Leibnitz-rule}), we arrive at
\begin{equation}
\left.\Delta m_{ij}^{(n,k,l)}\right|_{p=p_{0}, q=q_{0}}
=\left[ \sum^i_{\begin{subarray}{c} \mu=0,\\ \mu: even \end{subarray}}\frac{\mathcal{C}_{1,\mu}[\mathcal{Q}_{31}(p_0)]}{\mu !}\right]
\times
\left[ \sum^j_{\begin{subarray}{c} \nu=0,\\ \nu: even \end{subarray}}\frac{\mathcal{C}_{2,\nu}[\mathcal{Q}_{32}(q_0)] }{\nu !}\right] \left.m^{(n,k,l)}_{i-\mu,j-\nu}\right|_{p=p_{0},  q=q_{0}}.
\end{equation}
It is easy to see that the coefficients of the first term in above two summations are  $\mathcal{C}_{1,0}[\mathcal{Q}_{31}(p_0)]=\mathcal{Q}_{31}(p_0)$ and $\mathcal{C}_{2,0}[\mathcal{Q}_{32}(q_0)]=\mathcal{Q}_{32}(q_0)$.
Thus, according to the calculation in Ref.\cite{BYang2020DNLS}, one has the following matrix relation
\begin{equation}\label{matrix-relation}
\left(\left.\Delta m_{2i-1,2j-1}^{(n,k,l)}\right|_{p=p_{0},  q=q_{0}} \right)_{1\leq i, j\leq N}
=L \left(\left. m_{2i-1,2j-1}^{(n,k,l)}\right|_{p=p_{0},  q=q_{0}} \right)_{1\leq i, j\leq N} U,
\end{equation}
where $L$ and $U$ are following lower and  upper triangular matrices with diagonal entries as $\mathcal{Q}_{31}(p_0)$ and $\mathcal{Q}_{32}(q_0)$, respectively. 
\begin{eqnarray*}
&&
L=
\left(
  \begin{array}{cccc}
    \mathcal{Q}_{31}(p_0) & 0 & \cdots & 0 \\
    \frac{\mathcal{C}_{1,2}[\mathcal{Q}_{31}(p_0)]}{2!} & \mathcal{Q}_{31}(p_0) & \cdots & 0 \\
    \vdots & \vdots & \ddots & \vdots \\
    \frac{\mathcal{C}_{1,2N-2}[\mathcal{Q}_{31}(p_0)]}{(2N-2)!} & \frac{\mathcal{C}_{1,2N-4}[\mathcal{Q}_{31}(p_0)]}{(2N-4)!} & \cdots & \mathcal{Q}_{31}(p_0) \\
  \end{array}
\right),
\\
&&
U=
\left(
  \begin{array}{cccc}
    \mathcal{Q}_{32}(q_0) & \frac{\mathcal{C}_{2,2}[\mathcal{Q}_{32}(q_0)] }{2 !} & \cdots & \frac{\mathcal{C}_{2,2N-2}[\mathcal{Q}_{32}(q_0)] }{(2N-2)!} \\
    0 & \mathcal{Q}_{32}(q_0) & \cdots & \frac{\mathcal{C}_{2,2N-4}[\mathcal{Q}_{32}(q_0)] }{(2N-4)!} \\
    \vdots & \vdots & \ddots & \vdots \\
    0 & 0 & \cdots & \mathcal{Q}_{32}(q_0) \\
  \end{array}
\right).
\end{eqnarray*}
Taking determinant on both sides of equation (\ref{matrix-relation}), we have 
\begin{equation}
\Delta \tau_{n,k,l}= \left[\mathcal{Q}_{31}(p_0) \mathcal{Q}_{32}(q_0)\right]^N \tau_{n,k,l} = \Omega^N_0 \tau_{n,k,l},
\end{equation}
with $\Omega_0=-\frac{p_0(p_0-a)}{q_0(q_0+a)}=-\frac{[\sqrt{b(b-a)}+b][\sqrt{b(b-a)}+b-a]}{[\sqrt{b(b-a)}-b][\sqrt{b(b-a)}-b+a]}$.
If we further restrict $a$ and $b$ to be pure imaginary parameters, $\Omega_0$ becomes a complex number whose modulus is one.
Hence, by setting $e^{{\rm i}\kappa}=\Omega^N_0$, the index reduction condition (\ref{index-reduction1}) is proved.

\subsection{Complex conjugacy condition} \label{sect-complexreduction}

As mentioned previously, by taking $a$ and $b$ as the following pure imaginary numbers
\begin{equation}\label{ab1}
a={\rm i}\alpha,\ \ b={\rm i}\alpha(1+\sigma\rho_1\rho_2),
\end{equation}
with the real constant $\alpha$, the roots (\ref{roots1}) can be rewritten as
\begin{equation}\label{roots2}
p_0=q^*_0=\sqrt{-\sigma\rho_1\rho_2\alpha^2(1+\sigma\rho_1\rho_2)}+{\rm i}\alpha(1+\sigma\rho_1\rho_2),
\end{equation}
which implies that the condition for the existence of rogue wave solution: $-1<\rho_1\rho_2<0$ for $\sigma=1$ or $0<\rho_1\rho_2<1$ for $\sigma=-1$ need to be satisfied.
Besides, by imposing the parameter constraint $\eta_0=\xi^*_0$ and noticing that the coordinate transformations $x_1=\frac{\rho_2}{\alpha\rho_1}x$ and $x_{-1}=-\frac{\rho_1 \alpha(1+\sigma\rho_1\rho_2)}{\rho_2}t$ are real,
one can find that
\begin{equation}
\left.\left[  m_{i,j}^{(n,k,l)} \right]^*\right|_{p=p_{0}, q=q_{0}}=\left.\left[  m_{j,i}^{(-n-1,-k,-l)} \right]\right|_{p=p_{0}, q=q_{0}},
\end{equation}
which implies $\tau^*_{n,k,l}=\tau_{-n-1,-k,-l}$. By setting $n=-1,k=0,l=0$, one has the complex conjugacy conditions
\begin{equation}
\tau_{-1,0,0}=\tau^*_{0,0,0},\ \ \tau_{0,-1,0}=\tau^*_{-1,1,0},\ \ \tau_{0,0,-1}=\tau^*_{-1,0,1}.
\end{equation}
Finally, by defining the following variable transformations
\begin{equation*}
\tau_{0,0,0,}=f, \ \tau_{-1,0,0}=f^*, \  \tau_{-1,1,0}=e^{{\rm i}\kappa_0}g, \  \tau_{0,-1,0}=e^{-{\rm i}\kappa_0}g^*, \  \tau_{-1,0,1}=h, \  \tau_{0,0,-1}=h^*,
\end{equation*}%
we arrive at exactly the bilinear equations (\ref{MTdkBL1})--(\ref{MTdkBL4}).

%

\subsection{Rogue wave solutions in differential operator form}

Finally, based on the technique in Ref.\cite{BYang2020DNLS,BYang2021TW}, one can introduce free parameters through the
arbitrary parameter $\xi_0$.
To be specific,  we take parameter $\xi_0$ in the form
\begin{equation}
\xi_0=\sum^{\infty}_{r=1}\hat{a}_r\ln^r\mathcal{W}_1(p)=\sum^{\infty}_{r=1}\hat{a}_r\ln^r\left(\frac{p-b}{\sqrt{b(b-a)}}\right),
\end{equation}
where $\hat{a}_r$ are arbitrary complex parameters.

To summarize the above results and take dummy variables $t_a$ and $t_b$ as zeros,
we have the following theorem for rogue wave solutions to the MT model (\ref{MTa})-(\ref{MTb}):

\begin{theorem}\label{theorem-operator}
The MT  model (\ref{MTa})--(\ref{MTb}) possesses the following rogue wave solutions
\begin{equation} \label{var_tran2}
u=\rho_1 \frac{g}{f^{\ast} } e^{\mathrm{i}(1+\sigma \rho_1\rho_2) \left(\frac{\rho_2}{\rho_1}x+\frac{\rho_1}{\rho_2}t\right)}\,, \quad
v=\rho_2 \frac{{h}}{{f}} e^{\mathrm{i}(1+\sigma \rho_1\rho_2) \left(\frac{\rho_2}{\rho_1}x+\frac{\rho_1}{\rho_2}t\right)}\,,
\end{equation}
where
\begin{equation}\label{taus-operator}
f=\tau_{0,0,0,},\ \ f^*=\tau_{-1,0,0}, \ \ g=\Omega^{-N}_0 \tau_{-1,1,0}, \  \  h= \tau_{-1,0,1},
\end{equation}
and the elements in the determinant $\tau_{n,k,l} = \det_{1\leq i, j\leq N}\left( \tilde{m}_{2i-1,2j-1}^{(n,k,l)} \right)$ are defined by
\begin{eqnarray*}
&&
\tilde{m}_{i,j}^{(n,k,l)}=\frac{\left[(p-b)\partial_{p}\right]^{i}}{ i !}\frac{\left[(q+b)\partial_{q}\right]^{j}}{ j !} \left.m^{(n,k,l)}\right|_{p=p_{0}, q=p^*_{0}},
\\
&&
m^{(n,k,l)}=\frac{\mathrm{i}p }{p +q }
\left( -\frac{p}{q}\right) ^{n}
\left( -\frac{p-a}{q+a}\right) ^{k}
\left( -\frac{p-b}{q+b} \right) ^{l}
e^{\Theta},
\\
&&
\Theta=\frac{{\rm i}\rho_2}{a\rho_1}(p+q)x + \frac{{\rm i}\rho_1 b}{\rho_2} \left(\frac{1}{p}+\frac{1}{q}\right) t + \sum^{\infty}_{r=1}\hat{a}_r\ln^r\left(\frac{p-b}{\sqrt{b(b-a)}}\right) +\sum^{\infty}_{r=1}\hat{a}^*_r\ln^r\left(\frac{q+b}{\sqrt{b(b-a)}}\right),
\end{eqnarray*}
with $\Omega_0=-\frac{p_0(p_0-{\rm i}\alpha)}{p^*_0(p^*_0+{\rm i}\alpha)}$,
$p_0=\sqrt{b(b-a)}+b$, $b=a(1+\sigma\rho_1\rho_2)$ and $a={\rm i}\alpha$. Here $\alpha$, $\rho_1$ and $\rho_2$ are arbitrary real parameters which satisfy the conditions: $-1<\rho_1\rho_2<0$ for $\sigma=1$ or $0<\rho_1\rho_2<1$ for $\sigma=-1$, and $a_r$ $(r=1,2,\cdots)$ are arbitrary complex parameters.
\end{theorem}

In the above theorem, the final rational solutions are independent of the parameter $\alpha$, since it does not appear in the background plane waves.
In fact, $\alpha$ can be removed by the appropriate scaling of $p$ and $q$.
More specifically, we assume $\alpha>0$ without loss of generality and reparameterize $p=\alpha \hat{p}$ and $q=\alpha \hat{p}$,
then one has the equivalent theorem:

\begin{theorem}\label{theorem-operator1}
The MT  model (\ref{MTa})--(\ref{MTb}) possesses the rogue wave solutions (\ref{var_tran2}) --(\ref{taus-operator}) with elements of the determinant $\tau_{n,k,l} = \det_{1\leq i, j\leq N}\left( \tilde{m}_{2i-1,2j-1}^{(n,k,l)} \right)$ are defined by
\begin{eqnarray*}
&&
\tilde{m}_{i,j}^{(n,k,l)}=\frac{\left[(\hat{p}-\hat{b})\partial_{\hat{p}}\right]^{i}}{ i !}\frac{\left[(\hat{q}+\hat{b})\partial_{\hat{q}}\right]^{j}}{ j !} \left.m^{(n,k,l)}\right|_{\hat{p}=\hat{p}_{0}, \hat{q}=\hat{p}^*_{0}},
\\
&&
m^{(n,k,l)}=\frac{\mathrm{i}\hat{p} }{\hat{p} +\hat{q} }
\left( -\frac{\hat{p}}{\hat{q}}\right) ^{n}
\left( -\frac{\hat{p}-{\rm i}}{\hat{q}+{\rm i}}\right) ^{k}
\left( -\frac{\hat{p}-\hat{b}}{\hat{q}+\hat{b}} \right) ^{l}
e^{\Theta},
\\
&&
\Theta=\frac{\rho_2}{\rho_1}(\hat{p}+\hat{q})x + \frac{{\rm i}\rho_1 \hat{b}}{\rho_2} \left(\frac{1}{\hat{p}}+\frac{1}{\hat{q}}\right) t + \sum^{\infty}_{r=1}\hat{a}_r\ln^r\left(\frac{\hat{p}-\hat{b}}{\sqrt{\hat{b}(\hat{b}-{\rm i})}}\right) +\sum^{\infty}_{r=1}\hat{a}^*_r\ln^r\left(\frac{\hat{q}+\hat{b}}{\sqrt{\hat{b}(\hat{b}-{\rm i})}}\right),
\end{eqnarray*}
with $\Omega_0=-\frac{\hat{p}_0(\hat{p}_0-{\rm i})}{\hat{p}^*_0(\hat{p}^*_0+{\rm i})}$,
$\hat{p}_0=\sqrt{\hat{b}(\hat{b}-{\rm i})}+\hat{b}$ and $\hat{b}={\rm i}(1+\sigma\rho_1\rho_2)$. Here $\rho_1$ and $\rho_2$ are arbitrary real parameters which need to satisfy the conditions: $-1<\rho_1\rho_2<0$ for $\sigma=1$ or $0<\rho_1\rho_2<1$ for $\sigma=-1$, and $a_r$ $(r=1,2,\cdots)$ are arbitrary complex parameters.
\end{theorem}

\subsection{Rogue wave solutions through Schur polynomials}

In this subsection, rogue wave solutions will be presented by elementary Schur polynomials.
Following the technique in Ref.\cite{BYang2021TW}, the extended generator $\mathcal{G}$ of differential operators $\left[f_1 \partial_{p}\right]^{i}\left[f_2 \partial_{q}\right]^{j}$ is introduced as
\begin{equation}\label{generator-G1}
\mathcal{G}=\sum^{\infty}_{i=0}\sum^{\infty}_{j=0}\frac{\kappa^i}{i!}\frac{\lambda^j}{j!}\left[f_1 \partial_{p}\right]^{i}\left[f_2 \partial_{q}\right]^{j}
\end{equation}
which can be rewritten as
\begin{equation}\label{generator-G2}
\mathcal{G}=\sum^{\infty}_{i=0}\sum^{\infty}_{j=0}\frac{\kappa^i}{i!}\frac{\lambda^j}{j!}
\left[f_1 \partial_{\ln\mathcal{W}_1}\right]^{i}\left[f_2 \partial_{\ln\mathcal{W}_2}\right]^{j}
=\exp(\kappa \partial_{\ln\mathcal{W}_1} + \lambda \partial_{\ln\mathcal{W}_2}),
\end{equation}
through the transformation (\ref{f12-exp}).
According to the formula in Ref.\cite{Ohta2012NLS}, it is shown that for a function $F(\mathcal{W}_1,\mathcal{W}_2)$,
the following identity
\begin{equation}\label{generator-G21}
\mathcal{G}F(\mathcal{W}_1,\mathcal{W}_2)=F(e^{\kappa}\mathcal{W}_1,e^{\lambda}\mathcal{W}_2)
\end{equation}
 holds.

In the MT model, $p$ and $q$ can be solved explicitly with respect to $\mathcal{W}_1$ and $\mathcal{W}_2$ from the relations (\ref{W12-exp})
\begin{equation}\label{pq-W12-exp}
p=p(\mathcal{W}_1)=\mathcal{W}_1 \sqrt{b(b-a)}+b, \ \ q=q(\mathcal{W}_2)=\mathcal{W}_2\sqrt{b(b-a)}-b,
\end{equation}
in which $\mathcal{W}_1=\mathcal{W}_2=1$ when $p=p_0$ and $q=q_0$.
Then, by applying the relation (\ref{generator-G21}) to $m^{(n,k,l)}$ at $p=p_0,q=q_0$, one has
\begin{eqnarray}
\nonumber
\left.\mathcal{G}m^{(n,k,l)}\right|_{p=p_0,q=q_0}&=&\frac{(-1)^{n+k+l}\mathrm{i}p(\kappa) }{p(\kappa) +q(\lambda) }
\left[ \frac{p(\kappa)}{q(\lambda)}\right]^{n}
\left[ \frac{p(\kappa)-a}{q(\lambda)+a}\right]^{k}
\left[ \frac{p(\kappa)-b}{q(\lambda)+b} \right]^{l}
\times
\\
&&
\exp\left\{\frac{{\rm i}\rho_2x}{a\rho_1}\left[p(\kappa)+q(\lambda)\right]  +\frac{{\rm i}\rho_1 bt}{\rho_2}\left[\frac{1}{p(\kappa)}+\frac{1}{q(\lambda)}\right] + \sum^{\infty}_{r=1}\left [\hat{a}_r\kappa^r + \hat{a}^*_r\lambda^r\right] \right\},
\end{eqnarray}
where
\begin{equation*}
p(\kappa)\equiv \left.p(\mathcal{W}_1)\right|_{\mathcal{W}_1=e^{\kappa}}=e^{\kappa} \sqrt{b(b-a)}+b,\ \
q(\lambda)\equiv \left.q(\mathcal{W}_2)\right|_{\mathcal{W}_2=e^{\lambda}}=e^{\lambda}\sqrt{b(b-a)}-b.
\end{equation*}
Since
\begin{eqnarray*}
&&
\left.m^{(n,k,l)}\right|_{p=p_0,q=q_0}=\frac{(-1)^{n+k+l}\mathrm{i}p_0 }{p_0 +q_0 }
\left( \frac{p_0}{q_0}\right) ^{n}
\left( \frac{p_0-a}{q_0+a}\right) ^{k}
\left[ \frac{p_0-b}{q_0+b} \right] ^{l}
 \exp\left[\frac{{\rm i}\rho_2x}{a\rho_1}(p_0+q_0) +\frac{{\rm i}\rho_1 bt}{\rho_2} \left(\frac{1}{p_0}+\frac{1}{q_0}\right) \right],
\end{eqnarray*}
we deduce
\begin{eqnarray}\label{Gmm}
\nonumber
&&
\left.\frac{\mathcal{G}m^{(n,k,l)}}{m^{(n,k,l)}}\right|_{p=p_0,q=q_0}
\\
\nonumber
&&\hspace{0.5cm}=\frac{p_0+q_0}{p(\kappa) +q(\lambda) }
\left[ \frac{p(\kappa)}{p_0}\right]^{n+1}
\left[ \frac{q(\kappa)}{q_0}\right]^{-n}
\left[ \frac{p(\kappa)-a}{p_0-a}\right]^{k}
\left[ \frac{q(\lambda)+a}{q_0+a}\right]^{-k}
\left[ \frac{p(\kappa)-b}{p_0-b} \right]^{l}
\left[ \frac{q(\lambda)+b}{q_0+b} \right]^{-l}
\\
&&\hspace{0.7cm}
\times \exp\left\{\frac{{\rm i}\rho_2x}{a\rho_1}\left[p(\kappa)-p_0+q(\lambda)-q_0\right]  +\frac{{\rm i}\rho_1 bt}{\rho_2}\left[\frac{1}{p(\kappa)}-\frac{1}{p_0}+\frac{1}{q(\lambda)}-\frac{1}{q_0}\right] + \sum^{\infty}_{r=1}\left [\hat{a}_r\kappa^r + \hat{a}^*_r\lambda^r\right] \right\}.\ \ \
\end{eqnarray}

Next, the r.h.s of the above equation needs to be expanded in terms of  power series of $\kappa$ and $\lambda$.
By means of the techniques in Ref.\cite{BYang2021TW}, the first term is expressed as
\begin{eqnarray}\label{first-term1}
&&
\frac{p_0+q_0}{p(\kappa)+q(\lambda)}
=\sum^{\infty}_{\gamma=0}
\left(\frac{p_1q_1}{(p_0+q_0)^2}\kappa\lambda \right)^{\gamma}
\exp\left(\sum^{\infty}_{r=1}(\gamma s_r-b_r)\kappa^r + (\gamma s^*_r-b^*_r)\lambda^r \right),
\end{eqnarray}
where $p_1=\left.\frac{dp(\kappa)}{d\kappa}\right|_{\kappa=0}=\sqrt{b(b-a)}$ and $q_1=\left.\frac{dq(\lambda)}{d\lambda}\right|_{\lambda=0}=\sqrt{b(b-a)}$.
The parameters $s_r$ and $b_r$ are the expansion coefficients of $\kappa^r$ and $\lambda^r$ as follows:
\begin{eqnarray*}
&&
\ln\left[ \frac{p_0+q_0}{p_1\kappa}\frac{p(\kappa)-p_0}{p(\kappa)+q_0} \right]
=\ln\left[ \frac{2}{\kappa}\frac{e^\kappa-1}{e^\kappa+1} \right]
=\sum^{\infty}_{r=1} s_r \kappa^r,
\ \
\ln\left[ \frac{p(\kappa)+q_0}{p_0+q_0} \right]
=\ln\left[ \frac{1}{2}(e^\kappa+1) \right]
=\sum^{\infty}_{r=1} b_r \kappa^r,
\\
&&
\ln\left[ \frac{p_0+q_0}{q_1\lambda}\frac{q(\lambda)-q_0}{q(\lambda)+q_0} \right]
=\ln\left[ \frac{2}{\lambda}\frac{e^\lambda-1}{e^\lambda+1} \right]
=\sum^{\infty}_{r=1} s^*_r \lambda^r,
\ \
\ln\left[ \frac{q(\lambda)+p_0}{p_0+q_0} \right]
=\ln\left[ \frac{1}{2}(e^\lambda+1) \right]
=\sum^{\infty}_{r=1} b^*_r \lambda^r.
\end{eqnarray*}
Hence, $s^*_r=s_r$, $b^*_r=b_r$ and the term (\ref{first-term1}) is simplified as
\begin{eqnarray}
&&
\frac{p_0+q_0}{p(\kappa)+q(\lambda)}
=\sum^{\infty}_{\gamma=0}
\left(\frac{\kappa\lambda}{4} \right)^{\gamma}
\exp\left(\sum^{\infty}_{r=1}(\gamma s_r-b_r)(\kappa^r +\lambda^r) \right).
\end{eqnarray}

On the other hand, noticing that the complex conjugate relation $q(\lambda)=p^*(\lambda)$ and
$b={\rm i}\alpha(1+\sigma\rho_1\rho_2)\equiv{\rm i}\alpha\rho$,
we have the following expansion expressions
\begin{eqnarray*}
&&
\frac{{\rm i}\rho_2}{a\rho_1}\left[p(\kappa)-p_0\right]
=\frac{\rho_2}{\rho_1}[\sqrt{-\sigma \rho_1\rho_2 \rho}(e^\kappa-1)]
=\sum^{\infty}_{r=1}\alpha_r\kappa^r,\
\\
&&
\frac{{\rm i}\rho_1 b}{\rho_2}\left[\frac{1}{p(\kappa)}-\frac{1}{p_0}\right]
=\frac{\rho_1}{\rho_2}\left[\frac{\rho}{\sqrt{-\sigma \rho_1\rho_2 \rho}+{\rm i}\rho}-\frac{\rho}{e^\kappa\sqrt{-\sigma \rho_1\rho_2 \rho}+{\rm i}\rho}\right]
=\sum^{\infty}_{r=1}\beta_r\kappa^r,
\\
&&
\ln\frac{p(\kappa)}{p_0}
=\ln\frac{e^\kappa\sqrt{-\sigma \rho_1\rho_2 \rho}+{\rm i}\rho}{\sqrt{-\sigma \rho_1\rho_2 \rho}+{\rm i}\rho}=\sum^{\infty}_{r=1}\theta_r\kappa^r,\ \
\\
&&
\ln\frac{p(\kappa)-a}{p_0-a}
=\ln\frac{e^\kappa\sqrt{-\sigma \rho_1\rho_2 \rho}+{\rm i}\sigma\rho_1\rho_2}{\sqrt{-\sigma \rho_1\rho_2 \rho}+{\rm i}\sigma\rho_1\rho_2}
=\sum^{\infty}_{r=1}\vartheta_r\kappa^r,\ \
\\
&&
\ln\frac{p(\kappa)-b}{p_0-b}=\ln e^\kappa =\sum^{\infty}_{r=1}\zeta_r\kappa^r.
\end{eqnarray*}
With the help of these expansions, the rest terms on the r.h.s. of (\ref{Gmm}) are rewritten as
\begin{eqnarray*}
\exp\left\{\sum^{\infty}_{r=1}[\alpha_r x+\beta_r t +(n+1)\theta_r +k \vartheta_r + l \zeta_r +\hat{a}_r]\kappa^r
+ \sum^{\infty}_{r=1}[\alpha_r x+\beta^*_r t -n\theta^*_r -k \vartheta^*_r -l \zeta_r +\hat{a}^*_r]\lambda^r  \right\}.
\end{eqnarray*}
Thus, Eq.(\ref{Gmm}) is rewritten as
\begin{eqnarray}\label{gm-exp1}
\left.\frac{1}{m^{(n,k,l)}}\mathcal{G}m^{(n,k,l)}\right|_{p=p_0,q=q_0}
=\sum^{\infty}_{\gamma=0}
\left(\frac{\kappa\lambda}{4} \right)^{\gamma}
\exp\left(\sum^{\infty}_{r=1}(x^+_r(n,k,l)+\gamma s_r)\kappa^r+\sum^{\infty}_{r=1}(x^-_r(n,k,l)+\gamma s_r)\lambda^r\right),
\end{eqnarray}
where $x^\pm_r(n,k,l)$ are defined as
\begin{eqnarray*}
&&
x^+_r(n,k,l)=\alpha_r x+\beta_r t +(n+1)\theta_r +k \vartheta_r + l \zeta_r +\hat{a}_r-b_r,
\\
&&
x^-_r(n,k,l)=\alpha_r x+\beta^*_r t -n\theta^*_r -k \vartheta^*_r -l \zeta_r +\hat{a}^*_r-b_r.
\end{eqnarray*}

Denoting
\begin{equation}
\nonumber a_r=\hat{a}_r-b_r+\frac{1}{2}\theta_r,\ \ a^*_r=\hat{a}^*_r-b_r+\frac{1}{2}\theta^*_r,
\end{equation}
then the variables $x^\pm_r(n,k,l)$ are reparameterized as
\begin{eqnarray*}
&&
x^+_r(n,k,l)=\alpha_r x+\beta_r t +(n+\frac{1}{2})\theta_r +k \vartheta_r + l \zeta_r +a_r,
\\
&&
x^-_r(n,k,l)=\alpha_r x+\beta^*_r t -(n+\frac{1}{2})\theta^*_r -k \vartheta^*_r -l \zeta_r +a^*_r.
\end{eqnarray*}

Taking the coefficients of $\kappa^i\lambda^j$ on both sides of Eq.(\ref{gm-exp1}) and using the definition of Schur polynomial, one has
\begin{equation}
\frac{\tilde{m}_{i,j}^{(n,k,l)}}{\left.m^{(n,k,l)}\right|_{p=p_0,q=q_0}}
=\sum^{\mbox{min}(i,j)}_{\gamma=0}
\frac{1}{4^{\gamma}}
S_{i-\gamma}(\mbox{\boldmath $x$}^+(n,k,l)+\gamma \mbox{\boldmath $s$} )S_{j-\gamma}(\mbox{\boldmath $x$}^-(n,k,l)+\gamma \mbox{\boldmath $s$} ),
\end{equation}
where $\tilde{m}_{i,j}^{(n,k,l)}$ is the matrix element given in Theorem \ref{theorem-operator}.
Finally, using the gauge freedom of tau function, we define the following determinant
\begin{equation}
\sigma_{n,k,l}=\frac{\tau_{n,k,l}}{ \left(\left.m^{(n,k,l)}\right|_{p=p_0,q=q_0}\right )^N},
\end{equation}
which is also a rational solution to the MT model as presented in Theorem \ref{theorem1}.
Note that the denominator on the r.h.s of above equation is a gauge factor, which contains the ratio constant
$\Omega^N_0$ in the index reduction.
Hence, the function $g$ is changed into $g=\sigma_{-1,1,0}$ when rogue wave solutions are expressed through elementary Schur polynomials.

Finally, we comment that the algebraic equations (\ref{Q1-eq1}) actually allow another set of simple roots $(-p^*_0,-q^*_0)$, but it leads to the equivalent rogue wave solutions in the MT model (\ref{MTa})-(\ref{MTb}).
Indeed, one can find that when $p_0\rightarrow -p^*_0$, then $p(\kappa) \rightarrow -p^*(\kappa)$, $p_1\rightarrow -p_1$ and the expansion coefficients are changed into 
\begin{equation*}
s_r \rightarrow s_r, \ \ \alpha_r \rightarrow -\alpha_r, \ \ \beta_r \rightarrow -\beta^*_r,\ \  \theta_r \rightarrow \theta^*_r,\ \
\vartheta_r \rightarrow -\vartheta^*_r,\ \ \zeta_r \rightarrow \zeta_r.
\end{equation*}
Without loss of generality, the free parameters $a_r$ can be rewritten as $a^*_r$,
it follows that in Theorem \ref{theorem-operator}
\begin{equation}
[x^{\pm}_r(n,k,l)](x,t)\rightarrow [x^{\pm}_r(n,k,l)]^*(-x,-t),\ \ m^{(n,k,l)}_{i,j}(x,t)\rightarrow [m^{(n,k,l)}_{i,j}]^*(-x,-t),
\end{equation}
which imply $\sigma_{n,k,l}(x,t) \rightarrow \sigma^*_{n,k,l}(-x,-t)$.
Consider the invariance of plane waves $F(x,t)\rightarrow F^*(-x,-t)$ in the variable transformations (\ref{var_tran1}),
then
\begin{equation}\label{changeuv}
u(x,t)\rightarrow u^*(-x,-t),\ \ v(x,t)\rightarrow v^*(-x,-t),
\end{equation}
also satisfy the MT model (\ref{MTa})-(\ref{MTb}) in Theorem \ref{theorem-operator}.
However, the MT model (\ref{MTa})-(\ref{MTb}) is invariant under the variable transformation (\ref{changeuv}).
Hence, another group of roots $(-p^*_0,-q^*_0)$ gives rise to the equivalent rogue wave solutions through the appropriate parameter connections. This property is similar to that in the three-wave interaction system \cite{BYang2021TW}.

\section{Dynamics of rogue wave solutions}\label{sect-dynamics}

\subsection{First-order (fundamental) rogue wave solution}\label{sect-fisrtdynamics}

\begin{figure}[!htbp]
\centering
{\includegraphics[height=1.6in,width=4.0in]{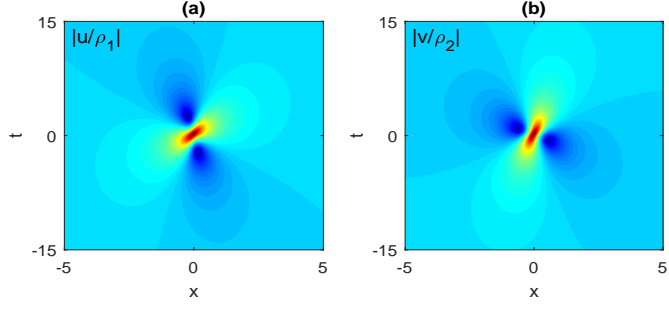}}
\caption{First-order rogue waves with the parameters $\sigma=-1$ and $\rho_1=0.5$.
\label{fig1}}
\end{figure}

\begin{figure}[!htbp]
\centering
{\includegraphics[height=1.6in,width=4.0in]{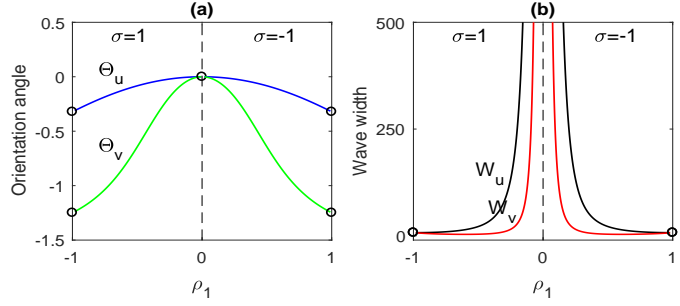}}
\caption{The orientation angles and wave widths of first-order rogue waves with respect to the varying parameter $\rho_1$.
\label{fig2}}
\end{figure}

According to Theorem \ref{theorem1}, when $N=1$, we obtain the first-order rogue wave solution as
\begin{eqnarray}\label{first-uv}
&&
u=\rho_1 e^{\mathrm{i}\phi_0} \left[1-\frac{d^*_2(L_1-d_1)-d_2(L^*_1+d^*_1)+|d_2|^2}{(L_1-d_1)(L^*_1+d^*_1) +\frac{d^2_0}{4} } \right],
\\
&&
v=\rho_2 e^{\mathrm{i}\phi_0} \left[1+\frac{(2d^*_1-d_0)(L_1+d_1)-(2d_1-d_0)(L^*_1-d^*_1)-|2d_1-d_0|^2}{(L_1+d_1)(L^*_1-d^*_1) +\frac{d^2_0}{4} } \right],
\end{eqnarray}
with
\begin{eqnarray*}
&&
L_1=\frac{\rho_2}{\rho_1}x+\frac{4\rho\rho_1d^2_1}{\rho_2}t,\ \
d_1=\frac{1}{2(\hat{\rho}+{\rm i}\rho)},\ \
d_2=\frac{1}{(\hat{\rho}+{\rm i}\sigma\rho_1\rho_2)},\ \ d_0=\frac{1}{\hat{\rho}},
\\
&&
\phi_0=\rho \left(\frac{\rho_2}{\rho_1}x+\frac{\rho_1}{\rho_2}t\right),\ \
 \hat{\rho}=\sqrt{-\sigma\rho_1\rho_2\rho},\ \ \rho=1+\sigma \rho_1\rho_2.
\end{eqnarray*}
It is quite clear that the structure of fundamental rogue wave is only determined by the background parameters $\rho_1$ and $\rho_2$ with the condition $\sigma \rho_1\rho_2(1+\sigma \rho_1\rho_2)<0$.
The intensities $|u|$ and $|v|$ tend to the backgrounds $|\rho_1|$ and $|\rho_2|$ as $x,t\rightarrow\infty$.
Through the local analyses, we can find that $|u|^2$ possesses one local maximum at $(x_1,t_1)$ with the peak amplitude $9\rho^2_1$ and two local minima at characteristic points $(x^{\pm}_2,t^{\pm}_2)$ with the zero-amplitude, while $|v|^2$ has one local maximum at $(x_1,t_1)$ with the peak amplitude $9\rho^2_2$ and two local minima at characteristic points $(x^{\pm}_3,t^{\pm}_3)$ with the zero-amplitude. Here three critical points are exactly defined by
\begin{equation}\label{criticalpoints}
(x_1,t_1)=(0,0), \ \
(x^{\pm}_2,t^{\pm}_2)=\left(\pm\frac{\Delta_1}{\rho_2}, \mp\frac{3\rho_2\Delta_1}{\rho^2_1} \right),\ \
(x^{\pm}_3,t^{\pm}_3)=\left(\mp\frac{3\rho_1\Delta_2}{\rho^2_2}, \pm\frac{\Delta_2}{\rho_1} \right),
\end{equation}
with $\Delta_1=\sqrt{\frac{-3\rho_1}{16\rho_2(3\rho_1\rho_2+4\sigma)}} $ and $\Delta_2=\sqrt{\frac{-3\rho_2}{16\rho_1(3\rho_1\rho_2+4\sigma)}}$.
Hence, the ratios between the peak and background values for two components are equal to 3.
For the rogue waves of the components $u$ and $v$, the orientation angles are $\Theta_u=\arctan\frac{-\rho^2_1}{3\rho^2_2}$ and $\Theta_v=\arctan\frac{-3\rho^2_1}{\rho^2_2}$, and the wave widths (i.e., the distance between two minima points) are $W_u=-\frac{3(\rho^4_1+9\rho^4_2)}{4\rho^3_1\rho^3_2(3\rho_1\rho_2+4\sigma)}$ and $W_v=-\frac{3(9\rho^4_1+\rho^4_2)}{4\rho^3_1\rho^3_2(3\rho_1\rho_2+4\sigma)}$, which indicate durations of the rogue waves.
Thus, the background parameters $\rho_1$ and $\rho_2$ do not affect the height of peak, but influence the orientation and duration
of two Peregrine solitons, which are observed graphically with $\sigma=1$ \cite{ye2021super}.
Besides, the appropriate scaling transformations (\ref{scaling-guo}) may bring the parameter $\mu$, which is physically important in the relativistic field theory and optics.
Then it is obvious that the orientation angles and wave widths depend on $\mu$, which naturally results in the rotation and lengthening of the fundamental rogue wave.
To illustrate the local structure of this rogue waves, we normalize $\rho_2=1$ and then $-1<\sigma\rho_1<0$.
Figure \ref{fig1} exhibits the amplitude profiles of two rogue waves, and Figures \ref{fig2} (a) and (b) display the orientation angles and wave widths with respect to $\rho_1$, respectively.

It is noted that the rogue waves in the MT model are different from the ones in the DNLS equation in spite of the fact that both systems belong to the same integrable hierarchy. 
For the rogue wave solutions of the DNLS equation, an internal parameter $\alpha$ exists and affects the structure of rogue wave dramatically \cite{BYang2020DNLS}.
In the case of the MT model, the same parameter $\alpha$ is introduced in the process of constructing rogue wave solutions as shown in Section \ref{sect-complexreduction}, but this parameter is removed finally and only two free parameters $\rho_1$ and $\rho_2$ control the shape of rogue wave.

In addition, unlike other integrable coupled systems, the MT model involving two components only supports the bright-type rogue wave.
In general, due to the additional degrees of freedom in coupled systems, there are more abundant patterns of rogue wave such as dark and four-petaled flower structures.
For the coupled MT model, the fundamental rogue wave solution does not contain extra parameters which can change the numbers of critical point in intensities, thus only allows the bright-type rogue wave.

\subsection{High-order rogue wave solutions}

The second-order rogue wave solution is obtained from Theorem \ref{theorem1} with $N=2$.
In this case, the tau functions $f$, $g$ and $h$ are given by
\begin{eqnarray}
f=\left\vert
\begin{array}{cc}
m^{(0,0,0)}_{11} & m^{(0,0,0)}_{13} \\
m^{(0,0,0)}_{31} & m^{(0,0,0)}_{33}
\end{array}\right\vert,
\ \
g=\left\vert
\begin{array}{cc}
m^{(-1,1,0)}_{11} & m^{(-1,1,0)}_{13} \\
m^{(-1,1,0)}_{31} & m^{(-1,1,0)}_{33}
\end{array}\right\vert,
\ \
h=\left\vert
\begin{array}{cc}
m^{(-1,0,1)}_{11} & m^{(-1,0,1)}_{13} \\
m^{(-1,0,1)}_{31} & m^{(-1,0,1)}_{33}
\end{array}\right\vert,
\end{eqnarray}
where the elements are determined by
\begin{eqnarray*}
&&
m^{(n,k,l)}_{11}=x^{+}_1\hat{x}^{-}_1+\frac{1}{4},
\ \
m^{(n,k,l)}_{13}=x^{+}_1\hat{x}^{-}_1+\frac{1}{8}\hat{x}^{-}_2,
\ \
m^{(n,k,l)}_{31}=x^{-}_1\hat{x}^{+}_1+\frac{1}{8}\hat{x}^{+}_2,
\\
&&
m^{(n,k,l)}_{33}=\hat{x}^{+}_1\hat{x}^{-}_1
+\frac{1}{16}\hat{x}^{+}_2\hat{x}^{-}_2
+\frac{1}{16}\left(x^{+}_1+2s_1\right)\left(x^{-}_1+2s_1\right)+\frac{1}{64},
\\
&&
\hat{x}^{+}_1=\frac{1}{6}(x^{+}_1)^3+x^{+}_1x^{+}_2+x^{+}_3,\ \ \hat{x}^{-}_1\hat{x}^{-}_1=\frac{1}{6}(x^{-}_1)^3+x^{-}_1x^{-}_2+x^{-}_3,
\\
&&
\hat{x}^{+}_2=\left(x^{+}_1+s_1\right)^2+2(s_2+x^{+}_2),\ \ \hat{x}^{-}_2=\left(x^{-}_1+s_1\right)^2+2(s_2+x^{-}_2),
\end{eqnarray*}
with the following coefficients:
\begin{eqnarray*}
&& a_1=a_2=0, \ \
\alpha_1=\frac{\rho_2\hat{\rho}}{\rho_1},\ \
\alpha_2=\frac{\rho_2\hat{\rho}}{2\rho_1},\ \
\alpha_3=\frac{\rho_2\hat{\rho}}{6\rho_1},\ \
\\
&&
\beta_1=\frac{\rho_1\rho\hat{\rho}}{\rho_2(\hat{\rho}+{\rm i}\rho)^2},\ \
\beta_2=\frac{\rho_1\rho^2(\sigma\rho_1\rho_2+{\rm i}\hat{\rho})}{2\rho_2(\hat{\rho}+{\rm i}\rho)^3},\ \
\beta_3=\frac{\rho_1\rho(4{\rm i}\sigma \rho_1\rho_2\rho^2-\hat{\rho}\rho^2+\hat{\rho}^{3})}{6\rho_2(\hat{\rho}+{\rm i}\rho)^4},
\\
&&
\theta_1=\frac{\hat{\rho}}{\hat{\rho}+{\rm i}\rho},\ \
\theta_2=\frac{{\rm i}\hat{\rho}\rho}{2(\hat{\rho}+{\rm i}\rho)^2},\ \
\theta_3=\frac{{\rm i}\rho^2(\rho-1+{\rm i}\hat{\rho})}{6(\hat{\rho}+{\rm i}\rho)^3},
\\
&&
\vartheta_1=\frac{\hat{\rho}}{{\rm i}(\rho-1)+\hat{\rho}},\ \
\vartheta_2=\frac{{\rm i}\hat{\rho}(\rho-1)}{2[{\rm i}(\rho-1)+\hat{\rho}]^2},\ \
\vartheta_3=\frac{({\rm i}\rho-\hat{\rho})(\rho-1)^2}{6[{\rm i}(\rho-1)+\hat{\rho}]^3},
\\
&&
\zeta_1=1,\ \ \zeta_2=\zeta_3=s_1=s_3=0,\ \ s_2=-\frac{1}{12},\ \ \hat{\rho}=\sqrt{-\sigma\rho_1\rho_2\rho},\ \ \rho=1+\sigma \rho_1\rho_2.
\end{eqnarray*}

\begin{figure}[!htbp]
\centering
{\includegraphics[height=4.0in,width=6.4in]{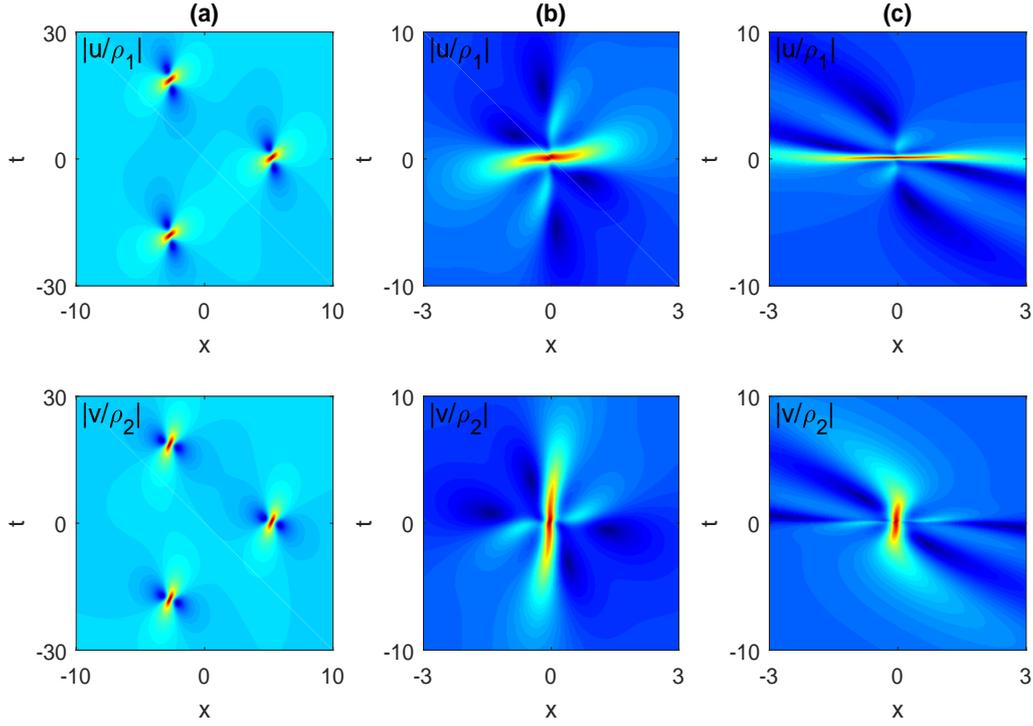}}
\caption{Second-order rogue waves with the parameters $\sigma=-1$. (a) a triangle pattern with $a_3=50$ and $\rho_1=0.5$; (b) a super rogue wave with $a_3=0$ and $\rho_1=0.5$; (c) a super rogue wave with $a_3=0$ and $\rho_1=0.9$.
\label{fig3}}
\end{figure}

From these explicit expressions, each tau function is a polynomial of degree six with respect to variables $x$ and $t$.
Three groups of the second-order rogue waves with different values of parameters $a_3$ and $\rho_1$ are illustrated in Figure \ref{fig3}.
The first group [Figure \ref{fig3} (a)] exhibits that a second-order rogue wave consists of three separating fundamental rogue waves and their distribution is a triangle array observably.
In the second group [Figure \ref{fig3} (b)], the parameter $\rho_1$ takes the same value as the one in the first group except $a_3=0$, thus,
it describes the parity-time-symmetric or super rogue wave in which three separate peaks come together and form a sole huge peak with a relative amplitude of $5$.
The last group [Figure \ref{fig3} (c)] also displays the super rogue wave except for a different value of $\rho_1$.
It is seen that the parameter $a_3$ controls the arrangement pattern of three individual first-order rogue waves, while the parameters $\rho_1$ and $\rho_2$ determine the orientations and durations of these rogue waves.

\begin{figure}[!htbp]
\centering
{\includegraphics[height=4in,width=6.4in]{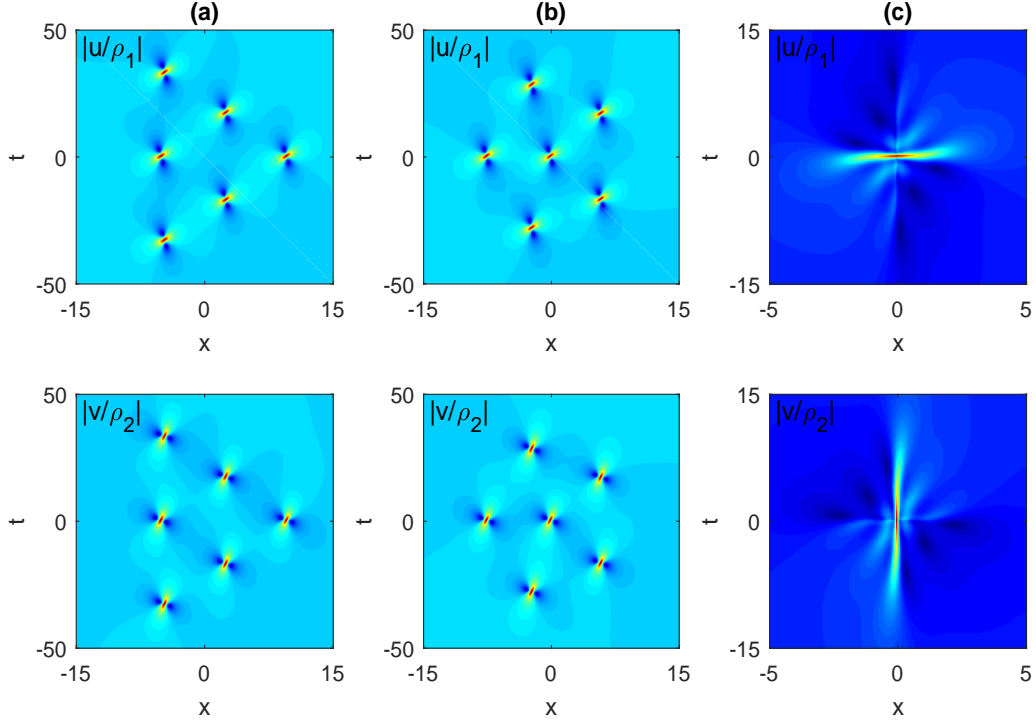}}
\caption{Third-order rogue waves with the parameters $\sigma=-1$ and $\rho_1=0.5$. (a) a triangle pattern with $(a_3,a_5)=(50,0)$; (b) a pentagon pattern with $(a_3,a_5)=(0,500)$; (c) a super rogue wave with $(a_3,a_5)=(0,0)$.
\label{fig4}}
\end{figure}

\begin{figure}[!htbp]
\centering
{\includegraphics[height=3in,width=6.8in]{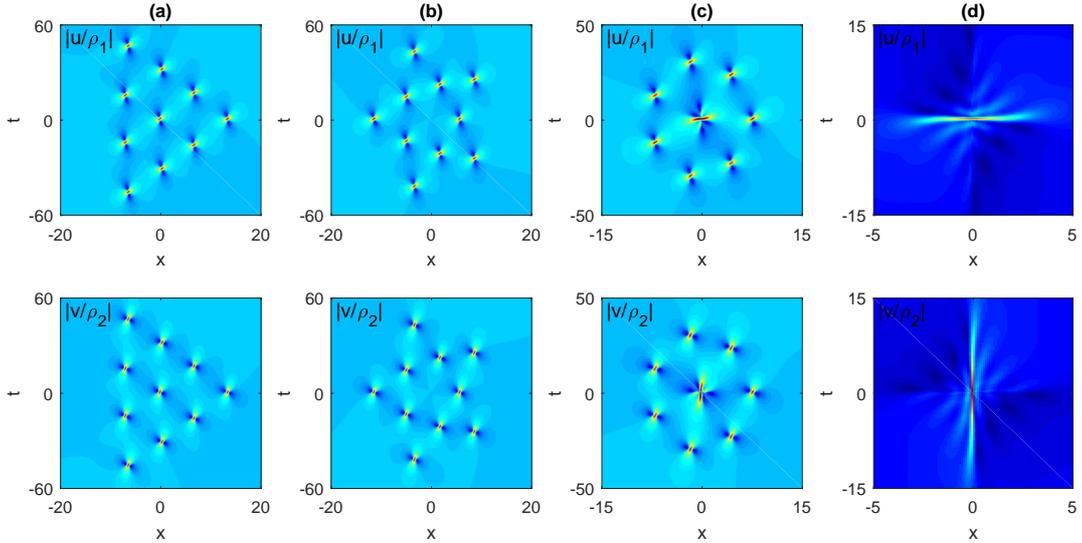}}
\caption{Fourth-order rogue waves with the parameters $\sigma=-1$ and $\rho_1=0.5$. (a) a triangle pattern with $(a_3,a_5,a_7)=(50,0,0)$; (b) a pentagon pattern with $(a_3,a_5,a_7)=(0,500,0)$; (c) a heptagon pattern with $(a_3,a_5,a_7)=(0,0,1000)$; (d) a super rogue wave with $(a_3,a_5,a_7)=(0,0,0)$.
\label{fig5}}
\end{figure}

In order to obtain third- and higher-order rogue waves, one need to take $N  \ge 3$ in Theorem \ref{theorem1}.
Thus, the corresponding tau functions are polynomials of the variables $x$ and $t$ with higher degree, which are too tedious to list here.
In Fig.\ref{fig4}, three groups of the third-order rogue waves ($N=3$) are presented graphically with a fixed value of $\rho_1$ but different values of $a_3$ and $a_5$.
One can observe that when the parameters value $(a_3,a_5)\neq(0,0)$, the third-order rogue wave exhibits the superposition of six fundamental ones and they obey different arrangements with  different values of  $a_3$ and $a_5$.
Specifically, the first group shows a triangle pattern with $(a_3,a_5)=(50,0)$ [Figure \ref{fig4} (a)], while the second group indicates a pentagon pattern with $(a_3,a_5)=(0,500)$ [Figure \ref{fig3} (b)].
In the last group [Figure \ref{fig4} (c)], the parameters value $(a_3,a_5)=(0,0)$ corresponds to the third-order super rogue wave
with a relative maximum amplitude of $7$. Furthermore, the 4th- and 5th-order rogue waves are illustrated in Figs.\ref{fig5} and \ref{fig6} with different parameters values.

From Figs.\ref{fig1}, \ref{fig3}-\ref{fig6} for the 1st to 5th-order
rogue waves, it can be concluded that the $N$th-order rogue wave in both components contains $N(N+1)/2$ fundamental ones in usual case $[(a_3,a_5,\cdots,a_{2N-1})\neq (0,0,\cdots,0)]$, while the relative amplitude of the huge peak is $2N+1$ in super case $[(a_3,a_5,\cdots,a_{2N-1})=(0,0,\cdots,0)]$.
The parameters $\rho_1$ and $\rho_2$ determine the orientation and duration of each rogue wave in both cases. However, the arrangement  of each individual rogue waves completely depends on the parameters $a_3,a_5,\cdots,a_{2N-1}$ in  the usual $N$th-order rogue wave.
In addition, it is found that the intensities of both components in the MT model possess the same relative amplitude of each peak in both the usual and super cases, especially the arrangement of fundamental rogue wave
is the same in the usual $N$th-order case.
The differences between the two components are the orientation and duration of each peak in either usual or super rogue wave.

\section{Rogue wave patterns with a single large internal parameter}
Rogue wave pattern refers to as the arrangement shape of fundamental rogue waves in a high-order case.
More recently, the universal rogue wave patterns have been studied by Yang and Yang \cite{BYang2021patterns1,BYang2021patterns2} in the integrable
systems such as the NLS equation, the derivative NLS equation, the Boussinesq equation and the Manakov system.
They predicted analytically that when one of the internal parameters in a higher order rogue wave is large, then this rogue wave exhibits certain geometric pattern with Peregrine waves forming shapes such as a triangle, pentagon, heptagon and so on with a possible lower-order rogue wave located at its center.
Through the asymptotical analysis, these patterns were shown to be associated with the root structure of the Yablonskii-Vorob'ev polynomial hierarchy under linear transformations.
In the MT model case, rogue wave patterns also agree with the universal laws existed in other integrable
systems \cite{BYang2021patterns1,BYang2021patterns2}, the analytical confirmations will be given in this section.
To this end, through the similar treatment as the NLS case that all $\Big{(}x_2^{\pm},x_4^{\pm},x_6^{\pm},\cdots \Big{)}$ are taken as zeros \cite{BYang2021patterns1}, we rewrite general rogue wave expressions in Theorem \ref{theorem1} as follows:

\begin{theorem}\label{theorem11}
The $N$-th order rogue waves in the MT model (\ref{MTa})-(\ref{MTb}) are given by
\begin{equation}\label{tans-uvN}
u_N(x,t)=\rho_1 \frac{g_N}{f^{\ast}_N } e^{\mathrm{i}(1+\sigma \rho_1\rho_2) \left(\frac{\rho_2}{\rho_1}x+\frac{\rho_1}{\rho_2}t\right)}\,, \quad
v_N(x,t)=\rho_2 \frac{h_N}{f_N} e^{\mathrm{i}(1+\sigma \rho_1\rho_2) \left(\frac{\rho_2}{\rho_1}x+\frac{\rho_1}{\rho_2}t\right)}\,,
\end{equation}
where
\begin{equation}
f_N=\sigma_{0,0,0,},\ \ f^*_N=\sigma_{-1,0,0}, \ \ g_N= \sigma_{-1,1,0}, \  \  h_N= \sigma_{-1,0,1},
\end{equation}
and the elements in determinant $\sigma_{n,k,l} = \det_{1\leq i, j\leq N}\left( m_{2i-1,2j-1}^{(n,k,l)} \right)$
\begin{equation}
m^{(n,k,l)}_{i,j}
=\sum^{\min (i,j)}_{\gamma=0}
\frac{1}{4^{\gamma}}
S_{i-\gamma}(\mbox{\boldmath $x$}^+(n,k,l)+\gamma \mbox{\boldmath $s$} )S_{j-\gamma}(\mbox{\boldmath $x$}^-(n,k,l)+\gamma \mbox{\boldmath $s$} ),
\end{equation}
with vectors $\mbox{\boldmath $x$}^{\pm}(n,k,l)=\Big{(}x_1^{\pm}(n,k,l),x_2^{\pm}(n,k,l),\cdots \Big{)}\equiv \Big{(}x_1^{\pm},x_2^{\pm},\cdots \Big{)}$ being defined by
\begin{eqnarray*}
&&
x^+_1 =\frac{\rho_2\hat{\rho}}{\rho_1}  x+\frac{\rho_1\rho\hat{\rho}}{\rho_2(\hat{\rho}+{\rm i}\rho)^2} t +(n+\frac{1}{2})\frac{\hat{\rho}}{\hat{\rho}+{\rm i}\rho} +k \frac{\hat{\rho}}{\hat{\rho}+{\rm i}\sigma\rho_1\rho_2} + l,
\\
&&
x^-_1 =\frac{\rho_2\hat{\rho}}{\rho_1}  x+\frac{\rho_1\rho\hat{\rho}}{\rho_2(\hat{\rho}-{\rm i}\rho)^2} t -(n+\frac{1}{2})\frac{\hat{\rho}}{\hat{\rho}-{\rm i}\rho} -k \frac{\hat{\rho}}{\hat{\rho}-{\rm i}\sigma\rho_1\rho_2} -l,
\\
&&
x^+_{2r+1} =\alpha_{2r+1} x+\beta_{2r+1} t +(n+\frac{1}{2})\theta_{2r+1} +k \vartheta_{2r+1}  +a_{2r+1},\ \ r\geq1,
\\
&&
x^-_{2r+1} =\alpha_{2r+1} x+\beta^*_{2r+1} t -(n+\frac{1}{2})\theta^*_{2r+1} -k \vartheta^*_{2r+1}  +a^*_{2r+1},\ \ r\geq1,
\\
&&
x^-_{2r}=0,\ \ r\geq1.
\end{eqnarray*}
These parameters $\alpha_r$, $\beta_r$, $\theta_r$, $\vartheta_r$ and $\mbox{\boldmath $s$}=\left(0,s_2,0, s_4\cdots\right)$ are coefficients from the following expansions
\begin{eqnarray*}
&&
\frac{\rho_2}{\rho_1}[\hat{\rho}(e^\kappa-1)]
=\sum^{\infty}_{r=1}\alpha_r\kappa^r,\
\ \
\frac{\rho_1\rho}{\rho_2}\left[\frac{1}{\hat{\rho}+{\rm i}\rho}-\frac{1}{e^\kappa\hat{\rho}+{\rm i}\rho}\right]
=\sum^{\infty}_{r=1}\beta_r\kappa^r,
\\
&&
\ln\frac{e^\kappa\hat{\rho}+{\rm i}\rho}{\hat{\rho}+{\rm i}\rho}=\sum^{\infty}_{r=1}\theta_r\kappa^r,
\ \
\ln\frac{e^\kappa\hat{\rho}+{\rm i}\sigma\rho_1\rho_2}{\hat{\rho}+{\rm i}\sigma\rho_1\rho_2}
=\sum^{\infty}_{r=1}\vartheta_r\kappa^r,\ \
\ \
\ln\left( \frac{2}{\kappa}\frac{e^\kappa-1}{e^\kappa+1} \right)
=\sum^{\infty}_{r=1} s_r \kappa^r,
\end{eqnarray*}
with $\hat{\rho}=\sqrt{-\sigma\rho_1\rho_2\rho}$ and $\rho=1+\sigma \rho_1\rho_2$.
Here $\rho_1$ and $\rho_2$ are arbitrary real parameters which satisfy the conditions: $-1<\rho_1\rho_2<0$ for $\sigma=1$ or $0<\rho_1\rho_2<1$ for $\sigma=-1$, and $a_3,a_5,\cdots,a_{2N-1}$ are arbitrary irreducible complex parameters.
\end{theorem}

\begin{figure}[!htbp]
\centering
{\includegraphics[height=2.5in,width=6.8in]{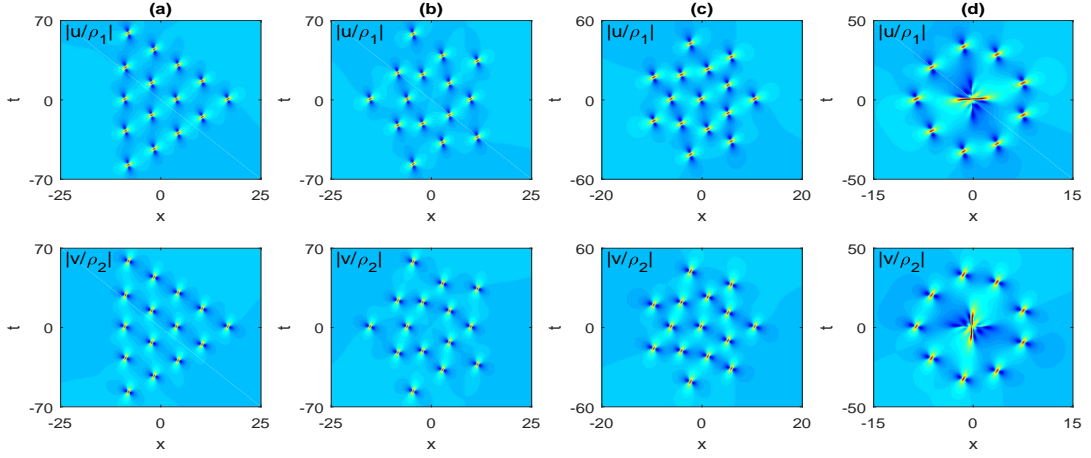}}
\caption{Fifth-order rogue waves with the parameters $\sigma=-1$ and $\rho_1=0.5$. (a) a triangle pattern with $(a_3,a_5,a_7,a_9)=(50,0,0,0)$; (b) a pentagon pattern with $(a_3,a_5,a_7,a_9)=(0,500,0,0)$; (c) a heptagon pattern with $(a_3,a_5,a_7,a_9)=(0,0,1000,0)$; (d) a enneagon pattern with $(a_3,a_5,a_7,a_9)=(0,0,0,2000)$; a super rogue wave with $(a_3,a_5,a_7,a_9)=(0,0,0,0)$.
\label{fig6}}
\end{figure}

In order to show the connection between the rogue wave pattern and root structure of the Yablonskii-Vorob'ev polynomials, we use the notations of the Yablonskii-Vorob'ev polynomial hierarchy in \cite{BYang2021patterns2}.
Then, the asymptotics of rogue waves with the large $a_{2m+1}$ in the MT model is summarized by the following theorem.

\begin{theorem}\label{theorem-patterns}
Let $[u_N(x,t),v_N(x,t)]$ be the $N$-th order rogue wave of the MT model in Eq.(\ref{tans-uvN}) where $a_{2m+1}$ is large and the other internal parameters are $O(1)$.

(1) Far away from the origin, with $\sqrt{x^2+t^2}=O(|a_{2m+1}|^{1/(2m+1)})$, this $[u_N(x,t),v_N(x,t)]$ asymptotically separates into $N_p$ fundamental rogue waves, where $N_p=\frac{1}{2}[N(N+1)-N_0(N_0+1)]$.
These fundamental rogue waves are $[\hat{u}_1(x-\hat{x}_0,t-\hat{t}_0)e^{\mathrm{i}(1+\sigma \rho_1\rho_2) \left(\frac{\rho_2}{\rho_1}x+\frac{\rho_1}{\rho_2}t\right)},\hat{v}_1(x-\hat{x}_0,t-\hat{t}_0)e^{\mathrm{i}(1+\sigma \rho_1\rho_2) \left(\frac{\rho_2}{\rho_1}x+\frac{\rho_1}{\rho_2}t\right)}]$, where
\begin{eqnarray}
&&\label{hatu1}
\hat{u}_1(x,t)=\rho_1
\frac{\hat{\rho}^2\left[\frac{\rho_2}{\rho_1}x + \frac{\rho_1\rho}{\rho_2(\hat{\rho}-{\rm i}\rho)^2}t +\frac{1}{2(\hat{\rho}-{\rm i}\rho)} -\frac{1}{\hat{\rho}-{\rm i}\sigma\rho_1\rho_2} \right]\left[\frac{\rho_2}{\rho_1}x + \frac{\rho_1\rho}{\rho_2(\hat{\rho}+{\rm i}\rho)^2}t -\frac{1}{2(\hat{\rho}+{\rm i}\rho)} +\frac{1}{\hat{\rho}+{\rm i}\sigma\rho_1\rho_2} \right]+\frac{1}{4}}{\hat{\rho}^2\left[\frac{\rho_2}{\rho_1}x + \frac{\rho_1\rho}{\rho_2(\hat{\rho}-{\rm i}\rho)^2}t +\frac{1}{2(\hat{\rho}-{\rm i}\rho)} \right]\left[\frac{\rho_2}{\rho_1}x + \frac{\rho_1\rho}{\rho_2(\hat{\rho}+{\rm i}\rho)^2}t -\frac{1}{2(\hat{\rho}+{\rm i}\rho)} \right]+\frac{1}{4}},
\\
&&\label{hatv1}
\hat{v}_1(x,t)=\rho_2
\frac{\hat{\rho}^2\left[\frac{\rho_2}{\rho_1}x + \frac{\rho_1\rho}{\rho_2(\hat{\rho}-{\rm i}\rho)^2}t +\frac{1}{2(\hat{\rho}-{\rm i}\rho)} -\frac{1}{\hat{\rho}} \right]\left[\frac{\rho_2}{\rho_1}x + \frac{\rho_1\rho}{\rho_2(\hat{\rho}+{\rm i}\rho)^2}t -\frac{1}{2(\hat{\rho}+{\rm i}\rho)} +\frac{1}{\hat{\rho}} \right]+\frac{1}{4}}{\hat{\rho}^2\left[\frac{\rho_2}{\rho_1}x + \frac{\rho_1\rho}{\rho_2(\hat{\rho}+{\rm i}\rho)^2}t +\frac{1}{2(\hat{\rho}+{\rm i}\rho)} \right]\left[\frac{\rho_2}{\rho_1}x + \frac{\rho_1\rho}{\rho_2(\hat{\rho}-{\rm i}\rho)^2}t -\frac{1}{2(\hat{\rho}-{\rm i}\rho)} \right]+\frac{1}{4}},
\end{eqnarray}
with $\hat{\rho}=\sqrt{-\sigma\rho_1\rho_2\rho}$ and $\rho=1+\sigma \rho_1\rho_2$.
The positions $(\hat{x}_0,\hat{t}_0)$ of these fundamental rogue waves are given by
\begin{eqnarray}
&&\label{positions1}
\hat{x}_0=-\frac{\hat{\rho}}{\sigma \rho\rho^2_2}\mathfrak{R}(z_0\Omega) + \frac{1+2\sigma\rho_1\rho_2}{2\sigma\rho\rho^2_2}\mathfrak{I}(z_0\Omega),\ \
\\
&&\label{positions2}
\hat{t}_0=\frac{1}{2\sigma\rho\rho^2_2}\mathfrak{I}(z_0\Omega),
\end{eqnarray}
with $z_0$ being any of the $N_p$ non-zero simple roots of $Q^{[m]}_N(z)$, and $(\mathfrak{R},\mathfrak{I})$ representing the real and imaginary parts of a complex number.
The error of this fundamental rogue wave approximation is $O(|a_{2m+1}|^{-1/(2m+1)})$.
Expressed mathematically, when $|a_{2m+1}|\gg 1$ and $[(x-\hat{x}_0)^2+(t-\hat{t}_0)^2]^{1/2}=O(1)$, we have the following solution asymptotics
\begin{eqnarray}
&&
u_N(x,t;a_3,a_5,\cdot,a_{2N-1})=\hat{u}_1(x-\hat{x}_0,t-\hat{t}_0)e^{\mathrm{i}(1+\sigma \rho_1\rho_2) \left(\frac{\rho_2}{\rho_1}x+\frac{\rho_1}{\rho_2}t\right)} + O(|a_{2m+1}|^{-1/(2m+1)}),
\\
&&
v_N(x,t;a_3,a_5,\cdot,a_{2N-1})=\hat{v}_1(x-\hat{x}_0,t-\hat{t}_0)e^{\mathrm{i}(1+\sigma \rho_1\rho_2) \left(\frac{\rho_2}{\rho_1}x+\frac{\rho_1}{\rho_2}t\right)} + O(|a_{2m+1}|^{-1/(2m+1)}).
\end{eqnarray}
When $(x,t)$ is not in the neighborhood of any of these $N_p$ fundamental waves, or $\sqrt{x^2+t^2}$ is larger than $O(|a_{2m+1}|^{-1/(2m+1)})$, then $[u_N(x,t),v_N(x,t)]$ asymptotically approaches the constant-amplitude background
$[\rho_1e^{\mathrm{i}(1+\sigma \rho_1\rho_2) \left(\frac{\rho_2}{\rho_1}x+\frac{\rho_1}{\rho_2}t\right)},$ \\ $ \rho_2e^{\mathrm{i}(1+\sigma \rho_1\rho_2) \left(\frac{\rho_2}{\rho_1}x+\frac{\rho_1}{\rho_2}t\right)}]$ as $|a_{2m+1}|\rightarrow \infty$.

(2) In the neighborhood of the origin, where $\sqrt{x^2+t^2}=O(1)$, $[u_N(x,t),v_N(x,t)]$ is asymptotically a lower $N_0$-th order rogue wave
$[u_{N_0}(x,t),v_{N_0}(x,t)]$, where $N_0$ is obtained from $(N,m)$ by the formula
\begin{eqnarray}\label{N0formula}
N_0=\left\{
      \begin{array}{ll}
        N\ \ {\rm mod}\ \ (2m+1), & {\rm if} \ \ 0\leq N\ \ {\rm mod}\ \ (2m+1)\leq m \\
        2m-[ N\ \ {\rm mod}\ \ (2m+1)], & {\rm if} \ \ N\ \ {\rm mod}\ \ (2m+1)>m.
      \end{array}
    \right.
\end{eqnarray}
with $0\leq N_0\leq N-2$, and $[u_{N_0}(x,t),v_{N_0}(x,t)]$ is given by Eq.(\ref{tans-uvN})
with its internal parameters $(a_3,a_5,\cdots,a_{2N_0-1})$ being the first $N_0-1$ values in the parameter set ${a_3,a_5,\cdots,a_{2N-1}}$
of the original rogue wave $[u_N(x,t),v_N(x,t)]$.
The error of this lower-order rogue wave approximation is $O(|a_{2m+1}|^{-1})$.
Expressed mathematically, when $|a_{2m+1}|\gg 1$ and $\sqrt{x^2+t^2}=O(1)$,
\begin{eqnarray}
&&
u_N(x,t;a_3,a_5,\cdot,a_{2N-1})=u_{N_0}(x,t;a_3,a_5,\cdot,a_{2N_0-1}) + O(|a_{2m+1}|^{-1}),
\\
&&
v_N(x,t;a_3,a_5,\cdot,a_{2N-1})=v_{N_0}(x,t;a_3,a_5,\cdot,a_{2N_0-1}) + O(|a_{2m+1}|^{-1}).
\end{eqnarray}
If $N_0=0$, then there will not be such a lower-order rogue wave in the neighborhood of the origin, and
$[u_{N}(x,t),v_{N}(x,t)]$ asymptotically approaches the constant-amplitude background $[\rho_1e^{\mathrm{i}(1+\sigma \rho_1\rho_2) \left(\frac{\rho_2}{\rho_1}x+\frac{\rho_1}{\rho_2}t\right)},\rho_2e^{\mathrm{i}(1+\sigma \rho_1\rho_2) \left(\frac{\rho_2}{\rho_1}x+\frac{\rho_1}{\rho_2}t\right)}]$ \\ as $|a_{2m+1}|\rightarrow \infty$.
\end{theorem}

\emph{Proof.}
As $|a_{2m+1}|\gg1$ and other parameters are of $O(1)$ in the MT rogue wave solution (\ref{tans-uvN}) at $(x,t)$ where $\sqrt{x^2+t^2}=O(|a_{2m+1}|^{1/(2m+1)})$, one has \cite{BYang2021patterns1}
\begin{equation}
S_j(\mbox{\boldmath $x$}^+(n,k,l)+\gamma \mbox{\boldmath $s$})=S_j(\mbox{$\mathbf{v}$})\left[1+ O\left(a_{2m+1}^{-1/(2m+1)}\right)\right],\ \
|a_{2m+1}|\gg1,
\end{equation}
where
\begin{equation}
\mathbf{v}=\left(\frac{\rho_2\hat{\rho}}{\rho_1}  x+\frac{\rho_1\rho\hat{\rho}}{\rho_2(\hat{\rho}+{\rm i}\rho)^2} t,0,\cdots,0,a_{2m+1},0,\cdots\right).
\end{equation}
From the definition of Schur polynomials, one can see that $S_j(\mbox{$\mathbf{v}$})$ is related to the polynomial $p^{[m]}_j(z)$ as
\begin{equation}
S_j(\mbox{$\mathbf{v}$})=\Omega^jp^{[m]}_j(z),
\end{equation}
where
\begin{equation}
z=\Omega^{-1}\left(\frac{\rho_2\hat{\rho}}{\rho_1}  x+\frac{\rho_1\rho\hat{\rho}}{\rho_2(\hat{\rho}+{\rm i}\rho)^2} t\right),\ \
\Omega=\left(-\frac{2m+1}{2^{2m}}a_{2m+1}\right)^{\frac{1}{2m+1}}.
\end{equation}

Similar relations can also be obtained for $S_j(\mbox{\boldmath $x$}^-(n,k,l)+\gamma \mbox{\boldmath $s$})$.
Using these formulae and the similar steps as in Ref. \cite{BYang2021patterns1}, we find that
\begin{equation}
\sigma_{n,k,l}\sim \chi_1 |a_{2m+1}|^{\frac{N(N+1)}{2m+1}}\left|Q^{[m]}_N(z) \right|^2,\ \ |a_{2m+1}|\gg1,
\end{equation}
where
\begin{equation*}
\chi_1=c^{-2}_N\left(\frac{1}{2} \right)^{N(N-1)}\left(\frac{2m+1}{2^{2m}} \right)^{\frac{N(N+1)}{2m+1}}.
\end{equation*}
Since $\chi_1$ is independent of $n$, $k$ and $l$, the above equation shows that for large $a_{2m+1}$,
$[u_N(x,t),v_N(x,t)]$ would be on the unit background except at or near $(x,t)$ locations $(\hat{x}_0,\hat{t}_0)$ where
\begin{equation}
z_0=\Omega^{-1}\left(\frac{\rho_2\hat{\rho}}{\rho_1}  \hat{x}_0+\frac{\rho_1\rho\hat{\rho}}{\rho_2(\hat{\rho}+{\rm i}\rho)^2} \hat{t}_0\right)
\end{equation}
is a root of the polynomial $Q^{[m]}_N(z)$.

Next, we show that in the neighborhood of each  $(\hat{x}_0,\hat{t}_0)$ location, i.e., $\left[(x-\hat{x}_0)^2+(t-\hat{t}_0)^2\right]^{1/2}=O(1)$, the rogue wave $[u_N(x,t),v_N(x,t)]$ approaches the fundamental rogue wave 
$[\hat{u}_1(x-\hat{x}_0,t-\hat{t}_0)e^{\mathrm{i}(1+\sigma \rho_1\rho_2) \left(\frac{\rho_2}{\rho_1}x+\frac{\rho_1}{\rho_2}t\right)},$ \\ $ \hat{v}_1(x-\hat{x}_0,t-\hat{t}_0)e^{\mathrm{i}(1+\sigma \rho_1\rho_2) \left(\frac{\rho_2}{\rho_1}x+\frac{\rho_1}{\rho_2}t\right)}]$
for large $a_{2m+1}$. To this end, we notice that when $(x,t)$ is in the neighborhood of $(\hat{x}_0,\hat{t}_0)$, we have a more refined asymptotics for $S_j(\mbox{\boldmath $x$}^+(n,k,l)+\gamma \mbox{\boldmath $s$})$ as \cite{BYang2021patterns1}
\begin{equation}
S_j(\mbox{\boldmath $x$}^+(n,k,l)+\gamma \mbox{\boldmath $s$})=S_j(\mbox{$\mathbf{\hat{v}}$})\left[1+ O\left(a_{2m+1}^{-2/(2m+1)}\right)\right],
\end{equation}
where
\begin{equation}
\mathbf{\hat{v}}=\left(x^{+}_1,0,\cdots,0,a_{2m+1},0,\cdots\right).
\end{equation}

The polynomial $S_j(\mbox{$\mathbf{\hat{v}}$})$ is related to $p^{[m]}_j(z)$ as $S_j(\mbox{$\mathbf{\hat{v}}$})=\Omega^j p^{[m]}_j(\Omega^{-1}x^+_1)$, then we get a refined asymptotics for $S_j(\mbox{\boldmath $x$}^+(n,k,l)+\gamma \mbox{\boldmath $s$})$ through polynomials $p^{[m]}_j(z)$. Similar refined asymptotics can also be obtained for $S_j(\mbox{\boldmath $x$}^-(n,k,l)+\gamma \mbox{\boldmath $s$})$.
Using these refined asymptotics of $S_j(\mbox{\boldmath $x$}^{\pm}(n,k,l)+\gamma \mbox{\boldmath $s$})$ and following the same steps as in Ref.\cite{BYang2021patterns1}, we find that
\begin{eqnarray}
\nonumber \sigma_{n,k,l}&=&\hat{\chi_1}\left| \left[Q^{[m]}_N\right]'(z_0) \right|^2 |a_{2m+1}|^{\frac{N(N+1)-2}{2m+1}}
\left[x^{+}_1(x-\hat{x}_0,t-\hat{t}_0;n,k,l)x^{-}_1(x-\hat{x}_0,t-\hat{t}_0;n,k,l)+\frac{1}{4} \right]
\\
&&
\times\left[1+O\left(a_{2m+1}^{-1/(2m+1)}\right)\right],
\end{eqnarray}
where $\hat{\chi}_1=\left[(2m+1)2^{-2m}\right]^{-2/(2m+1)}\chi_1$.
Finally, the simplicity of the nonzero roots in Yablonskii-Vorob'ev polynomials implies that $\left[Q^{[m]}_N\right]'(z_0)\neq0$.
This indicates that the above leading-order asymptotics for $\sigma_{n,k,l}(x,t)$ does not vanish.
Therefore, when $a_{2m+1}$ is large and $(x,t)$ in the neighborhood of $(\hat{x}_0,\hat{t}_0)$, we obtain
\begin{eqnarray}
&&
u_N(x,t)=\hat{u}_1(x-\hat{x}_0,t-\hat{t}_0)e^{\mathrm{i}(1+\sigma \rho_1\rho_2) \left(\frac{\rho_2}{\rho_1}x+\frac{\rho_1}{\rho_2}t\right)} + O\left(a_{2m+1}^{-1/(2m+1)}\right),
\\
&&
v_N(x,t)=\hat{v}_1(x-\hat{x}_0,t-\hat{t}_0)e^{\mathrm{i}(1+\sigma \rho_1\rho_2) \left(\frac{\rho_2}{\rho_1}x+\frac{\rho_1}{\rho_2}t\right)} + O\left(a_{2m+1}^{-1/(2m+1)}\right),
\end{eqnarray}
and the error of this prediction is $O\left(a_{2m+1}^{-1/(2m+1)}\right)$.

Regarding the proof for the rogue pattern near the origin, it is very similar to that for the NLS and GDNLS equation in Refs.\cite{BYang2021patterns1,BYang2021patterns2}. So it is omitted here.

\begin{figure}[!htbp]
\centering
{\includegraphics[height=1.2in,width=5.2in]{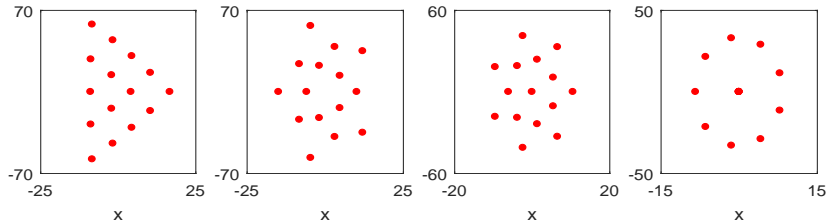}}
\caption{The locations $(\hat{x}_0,\hat{t}_0)$ associated with the root structures of $Q^{[m]}_5(z)$ with $1\leq m\leq4$ and the respective
large internal parameter $a_{2m+1}$ provided by Eq.(\ref{large-parameter}) and other system parameters are $\sigma=-1$, $2\rho_1=\rho_2=1$.
\label{fig7}}
\end{figure}

From Eqs.(\ref{positions1}) and (\ref{positions2}) in Theorem \ref{theorem-patterns}, one can see that the location $(\hat{x}_0,\hat{t}_0)$ of each fundamental rogue wave away from the center is given by the real and imaginary parts of each nonzero simple root $z_0$ of $Q^{[m]}_N(z)$ through the linear tansformation
\begin{equation}\label{linear-tansformation}
\left[
\begin{array}{c}
  \hat{x}_0 \\
  \hat{t}_0
\end{array}
\right]
=\left[
   \begin{array}{cc}
     -\frac{\hat{\rho}}{\sigma \rho\rho^2_2}\mathfrak{R}(\Omega) + \frac{1+2\sigma\rho_1\rho_2}{2\sigma\rho\rho^2_2}\mathfrak{I}(\Omega) & \frac{\hat{\rho}}{\sigma \rho\rho^2_2}\mathfrak{I}(\Omega) + \frac{1+2\sigma\rho_1\rho_2}{2\sigma\rho\rho^2_2}\mathfrak{R}(\Omega) \\
     \frac{1}{2\sigma\rho\rho^2_2}\mathfrak{I}(\Omega) & \frac{1}{2\sigma\rho\rho^2_2}\mathfrak{R}(\Omega) \\
   \end{array}
 \right]
\left[
\begin{array}{c}
  \mathfrak{R}(z_0) \\
  \mathfrak{I}(z_0)
\end{array}
\right].
\end{equation}
This linear transformation means that the whole rogue pattern formed by fundamental rogue waves in the $(x,t)$ plane is just a linear transformation matrix $B$ applied to the root structure of the Yablonskii-Vorob'ev polynomial $Q^{[m]}_N(z)$ in the complex $z$ plane.
The transformation matrix dictates the effects of stretch, shear and orientation to the root structure, which can be observed in Fig.\ref{fig7} that
exhibit the locations $(\hat{x}_0,\hat{t}_0)$ linked to the root structures of $Q^{[m]}_5(z)$ with $1\leq m\leq4$.
In Fig.\ref{fig7}, the parameters are chosen as the same values as that in Fig.\ref{fig6} (a)-(d) with $\sigma=-1$, $2\rho_1=\rho_2=1$ and
and the large internal parameter $a_{2m+1}$ respectively as
\begin{equation}\label{large-parameter}
(a_3,a_5,a_7,a_9)=(50,500,1000,2000).
\end{equation}

Next, we illustrate the analytical predictions of rogue wave patterns and compare them to true cases in the MT model.
According to Theorem \ref{theorem-patterns}, the prediction for rogue wave patterns can be assembled into the formulae
\begin{eqnarray}
&&
|u_{N}(x,t)|\approx|u_{N_0}(x,t)|+\sum^{N_p}_{k=1}\left(\left|\hat{u}_1(x-\hat{x}^{(k)}_0,t-\hat{t}^{(k)}_0)\right|-\rho_1 \right),
\\
&&
|v_{N}(x,t)|\approx|v_{N_0}(x,t)|+\sum^{N_p}_{k=1}\left(\left|\hat{v}_1(x-\hat{x}^{(k)}_0,t-\hat{t}^{(k)}_0)\right|-\rho_2 \right),
\end{eqnarray}
where $N_0$ is given by Eq.(\ref{N0formula}), $[u_{N_0}(x,t),v_{N_0}(x,t)]$ is the lower-order rogue wave at the center whose internal parameters $(a_3,a_5,\cdots,a_{2N_0-1})$ are inherited directly from those of the original rogue wave $[u_{N}(x,t),v_{N}(x,t)]$,
the function $[\hat{u}_1(x,t),\hat{v}_1(x,t)]$ is the fundamental rogue wave given in Eqs.(\ref{hatu1})-(\ref{hatv1}) with its position $(\hat{x}^{(k)}_0,\hat{t}^{(k)}_0)$ given by Eqs.(\ref{positions1})-(\ref{positions2}) for every one of the nonzero simple roots $z^{(k)}_0$ of $Q^{[m]}_N(z)$, and $N_p$ is the number of such fundamental rogue waves with $N_p=\frac{1}{2}[N(N+1)-N_0(N_0+1)]$.

\begin{figure}[!htbp]
\centering
{\includegraphics[height=3in,width=6.8in]{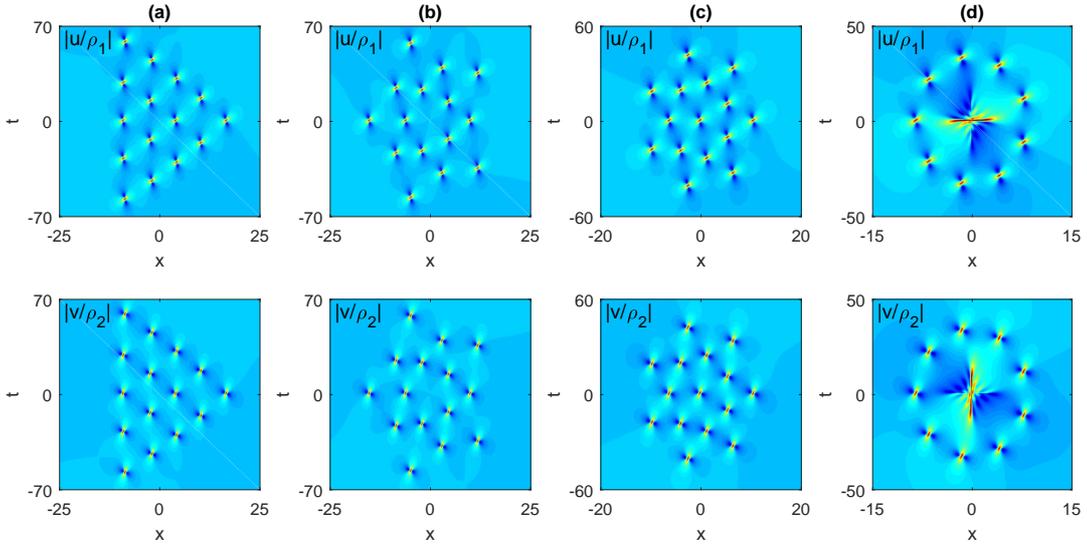}}
\caption{Predicted rogue wave patterns $|u_N(x,t)|$ and $|v_N(x,t)|$ for the order $N=5$ with the parameters $\sigma=-1$, $2\rho_1=\rho_2=1$.
The large internal parameters $a_{2m+1}$ are given in Eq.(\ref{large-parameter}) for (a)-(d) respectively, and the other parameters are taken as zero.
\label{fig8}}
\end{figure}

We only illustrate the  case for $N=5$ to show the comparison between these predicted rogue patterns and the true ones.
The parameters and large internal parameters are chosen as the same as those in Fig.\ref{fig6} (a)-(d) or in Fig.\ref{fig7}.
Under identical $(x,t)$ intervals, Fig.\ref{fig7} provides the predicted locations of fundamental rogue wave and lower-order rogue waves at the center in Fig.\ref{fig6} (a)-(d), and Fig.\ref{fig8} displays the complete predicted rogue wave patterns respectively.
It can be seen that the predicted rogue wave patterns coincide almost exactly with the true ones regarding the locations of
individual fundamental rogue waves, the overall shapes formed by these fundamental waves and the fine details of the lower-order
rogue waves at the center.
Theorem \ref{theorem-patterns} also states quantitatively the error of the predicted solution in two cases.
Similar to the NLS equation \cite{BYang2021patterns1}, the orders of these errors can be confirmed numerically and the details are omitted here.

\section{Summary and Discussions}

In this paper, we have constructed the general rogue wave solutions in the MT model by using the KP hierarchy reduction method.
These solutions are represented  in terms of determinants in which elements are given by elementary Schur-polynomials. In the process of constructing these solutions, we proved that two dimension-reduction conditions and one index-reduction condition are satisfied simulteneously under the same constraint of parameters.
The local structure analyses show that two background parameters influence the orientation and duration but they do not affect the heights of peak in rogue waves. 
It is also shown that compared with other integrable systems, there exist no additional parameters in rogue wave solutions of the MT model, and hence only bright-type rogue wave appears.
The $N$-the order rogue waves correspond to the superposition of $N(N+1)/2$ fundamental ones, and their arrangement patterns are determined by the values of non-reducible parameters.
By setting all internal parameters to be zero, we obtain the super rogue wave of $N$-th order in which the sole huge peak is located at the center and its maximum amplitude is  $2N+1$ times the background.
Finally, we have discussed rogue wave patterns when one of the internal parameters is large.
It is shown that similar to other integrable systems, these patterns are linked to the root structure of the Yablonskii-Vorob'ev polynomial hierarchy through a linear transformation.

\section*{Acknowledgement}
The work of J.C. is supported by the National Natural Science Foundation of China (No.11705077) and the Zhejiang Province Natural Science Foundation of China (Grant No. 2022SJGYZC01).
BF's work is
partially supported by National Science Foundation (NSF) under Grant No. DMS-1715991 and U.S. Department of Defense (DoD), Air Force for Scientific Research (AFOSR) under grant No. W911NF2010276.

\end{document}